\documentclass[aps,prb,twocolumn]{revtex4}

\usepackage{graphicx}
\usepackage{amsfonts}
\usepackage{amsmath,latexsym,amsthm}

\begin{document}

% \title{Surface gravity as a Noether charge}
\title{Noether charge astronomy}

\author{Zhi-Wei Wang${}^{1,2,\dagger}$
and Samuel L.\ Braunstein${}^{2,\ast}$}
\affiliation{${}^1$College of Physics, Jilin University,
Changchun, 130012, People's Republic of China}
\affiliation{${}^2$Computer Science, University of York, York YO10 5GH,
United Kingdom }
\affiliation{${}^\dagger$zhiweiwang.phy@gmail.com}
\affiliation{${}^\ast$sam.braunstein@york.ac.uk}

\begin{abstract}

\noindent Noether's theorem identifies fundamental conserved quantities,
called Noether charges, from a Hamiltonian. To-date Noether charges
remain largely elusive within theories of gravity: We do not know how to
directly measure them, and their physical interpretation remains
unsettled in general spacetimes. Here we show that the surface gravity
as naturally defined for a family of observers in arbitrarily dynamical
spacetimes is a directly measurable Noether charge. This Noether charge
reduces to the accepted value on stationary horizons, and, when
integrated over a closed surface, yields an energy with the
characteristics of gravitating mass. Stokes' theorem then identifies the
gravitating energy density as the time-component of a locally conserved
Noether current in general spacetimes. Our conclusion, that this Noether
charge is extractable from astronomical observations, holds the
potential for determining the detailed distribution of the gravitating
mass in galaxies, galaxy clusters and beyond.

\end{abstract}

\maketitle

In classical physics, Noether's theorem shows that if the evolution
generated by the Hamiltonian of a system leaves some quantity invariant
then conversely the evolution generated by that quantity will leave the
Hamiltonian invariant.\cite{Noether1918,Baez2018} Such conserved
quantities are called Noether charges and their conservation makes them
fundamental quantities of that system; notable examples include energy,
momentum and electric charge. The proof of Noether's result in the
Hamiltonian formulation is especially elegant\cite{Baez2018} (see
Appendix 1 for summary). Noether charges were first discovered in the
context of general relativity,\cite{Noether1918} and gravity is also the
focus of this paper.

Already in 1959, Komar\cite{Komar1959} found that
$J^\mu\equiv\xi^{[\nu;\mu]}_{~~~~~;\nu}$ is a locally conserved
4-current for an arbitrary vector field $\xi^\mu$. Indeed, its conservation
law $J^\mu_{~~;\mu} \equiv 0$, which holds for general dynamical 
spacetimes, follows purely from geometric considerations (see Appendix 1). 
Integrating the current $J^\mu$ on any hypersurface 
$\Sigma$ and applying Stokes' theorem yields
\begin{equation}
\frac{1}{4\pi}\!
\int_{\Sigma} J^\mu d\Sigma_\mu = 
\frac{1}{4\pi}\!
\int_{\partial\Sigma}
\xi^{[\mu;\nu]} d\Sigma_{\mu\nu},
\label{Stokes}
\end{equation}
where $d\Sigma_\mu$ and $d\Sigma_{\mu\nu}$ are the tensorial volume and
surface elements on the hypersurface $\Sigma$ and its boundary
$\partial\Sigma$, respectively. (See the Appendix 1 for a more detailed
discussion about the conservation laws associated with $J^\mu$.)
Thus, Eq.(\ref{Stokes}) immediately implies that
\begin{equation}
\xi^{[\mu;\nu]} d\Sigma_{\mu\nu}
\label{Q}
\end{equation}
is a local Noether charge, up to the addition of an exact 2-form, for
any theory of gravity with a torsion-free metric-compatible connection
(see Appendix 1). Using the Lagrangian formalism, (\ref{Q}) has been
confirmed as a Noether charge specifically for general
relativity\cite{Iyer1994,Bak1994} and additional Noether charges have
been identified for various diffeomorphically invariant theories of
gravity.\cite{Wald1993,Iyer1994}

% \newpage

One road block in the application of Komar's result has been in
identifying a choice of $\xi^\mu$ for which the Noether charge~(\ref{Q})
is directly observable.\cite{Fletcher1960} For stationary spacetimes,
the inherent anti-symmetry of the Killing condition makes the Killing
vector a natural choice of this vector field. Substituting
$\xi^\mu=\partial_t$ in Eq.~(\ref{Stokes}) leads to the Komar
mass.\cite{Komar1959} On asymptotically-flat stationary spacetimes the
Komar mass is equivalent to the ADM mass,\cite{Beig1978} and can be
interpreted as the force-per-unit-mass exerted at spatial infinity
needed to hold objects stationary. However, beyond the stationary
setting, the Komar mass has eluded interpretation.\cite{Bak1994}

Here we find a choice for $\xi^\mu$ even for arbitrarily dynamical
spacetimes, which yields a physically meaningful Noether charge. We
prove the utility of this choice of $\xi^\mu$ by showing: that~(\ref{Q})
reduces to the surface gravity experienced by a family of observers on
some surface of interest; that the corresponding integral,
Eq.~(\ref{Stokes}), corresponds to the gravitating mass; and that its
density, $J^\mu d\Sigma_\mu/4\pi$, is the gravitating energy density.
Finally, we illustrate how the detailed distribution of gravitating mass
in astronomical systems may be determined from red-shift factors.

\section*{Surface gravity as a Noether charge}

\begin{figure}[ht]
\centering
\vskip -0.1truein
\includegraphics[width=0.38\textwidth]{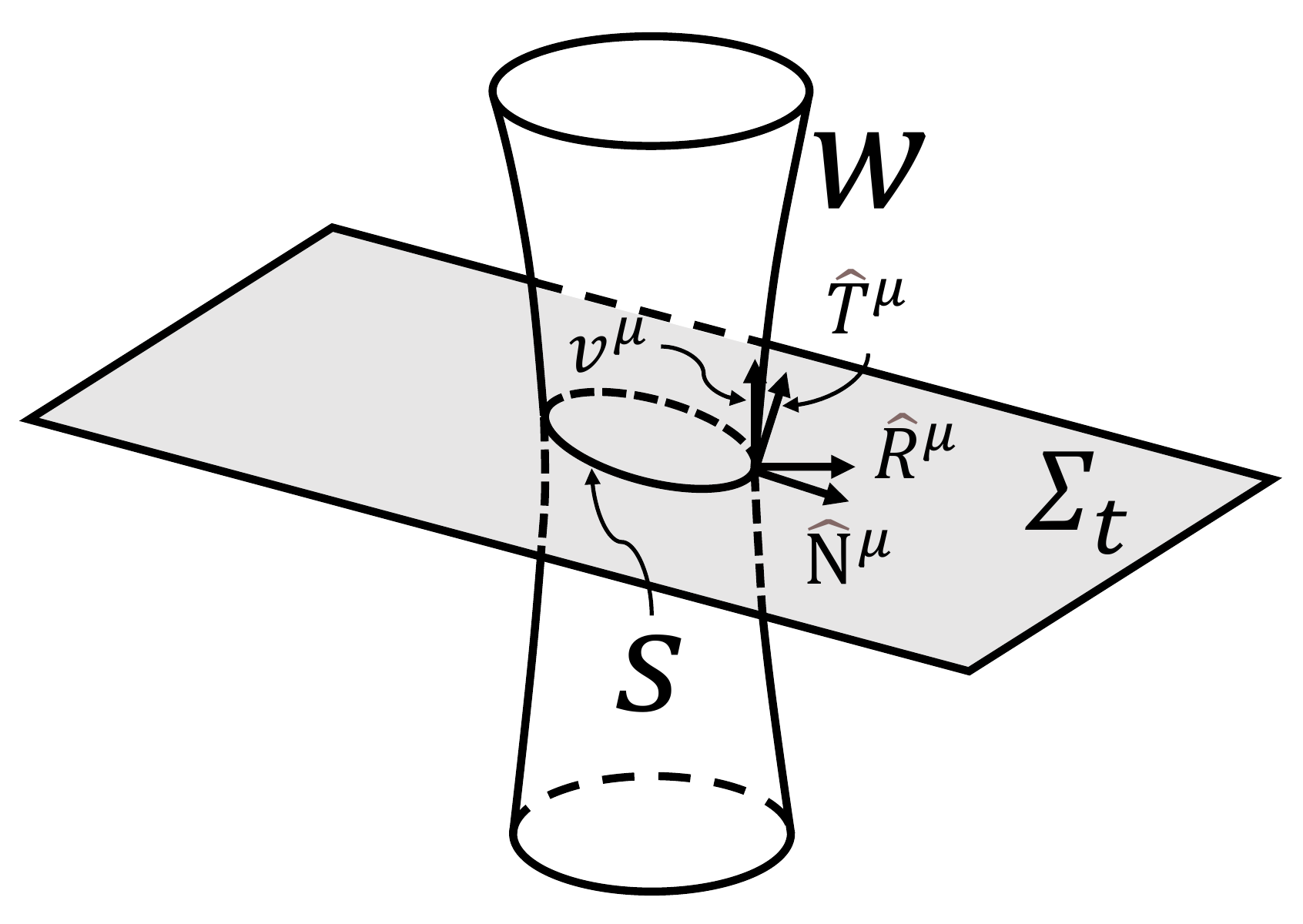}
\vskip -0.1truein
\caption{A timelike world tube ${\cal W}$ intersects a spacelike
hypersurface $\Sigma_t$ on the 2-surface ${\cal S}={\cal W}\cap
\Sigma_t$. An observer at some point on ${\cal S}$ will have: a
4-velocity $v^\mu$ that is tangent to ${\cal W}$ and orthogonal to the
tangent space of ${\cal S}$; an outgoing 4-vector $\hat R^\mu$ that is 
normal to ${\cal W}$; a future-directed timelike unit 4-vector $\hat
T^\mu$ normal to $\Sigma_t$; and an outgoing spacelike unit 4-vector
$\hat N^\mu$ orthogonal to the tangent space of ${\cal S}$ and tangent
to $\Sigma_t$.}
\vskip -0.1truein
\label{fig0}
\end{figure}

Consider a congruence of (in general non-geodesic) timelike observers
moving tangent to some world tube ${\cal W}$. We label the observers'
4-velocities as $v^\mu$ and their red-shift factors by $\Phi\equiv d\tau
/dt$, where $\tau$ is their proper time and $t$ is the coordinate time.
Throughout the paper we work in geometric units where $G=c=1$.
Although not necessary for this formalism, we envision $t$ as the
proper time of some, possibly distant, reference observer. For each
`time slice' there is a 2-surface of intersection ${\cal S}$ described
by the pointwise intersection of the world tube and hypersurfaces
$\Sigma_t$ of constant $t$, ${\cal S} ={\cal W}\cap \Sigma_t$. See
Fig.~\ref{fig0}. To measure a surface gravity at points on the 2-surface
${\cal S}$ we take our observers' 4-velocities $v^\mu$ to be orthogonal
to the tangent space of this surface.

Consider an observer from our congruence accelerating through some point
on the 2-surface ${\cal S}$. The surface gravity felt by this
observer, rescaled to the rate-of-change of coordinate time, is 
\begin{equation}
\Phi \,a_\mu \hat R^\mu,
\label{kappaN}
\end{equation}
where $a_\mu\equiv v_{\mu;\nu} v^\nu$ is the 4-acceleration of the
observer and $\hat R^\mu$ is normal to ${\cal W}$ at this point (see
Fig.~\ref{fig0}). Note, that $a_\mu \hat R^\mu$ is the component of
acceleration normal to the 2-surface in the observer's instantaneous
rest frame. We shall now show that (\ref{kappaN})
is a Noether charge in the following sense:

\vskip 0.1in

\noindent
{\bf Theorem:} Let $x^\mu(\tau)$ be the trajectory of an observer moving
orthogonally to the tangent space of an infinitesimal element of surface
area $dA$ of the 2-surface ${\cal S}$. Then choosing 
$\xi^\mu \equiv dx^\mu(\tau)/dt$ yields
$\kappa_{\text{Noether}}\, dA= \xi^{[\mu;\nu]} d\Sigma_{\mu\nu}$, for 
surface gravity $\kappa_{\text{Noether}} \equiv \Phi \,a_\mu \hat R^\mu$.

\vskip 0.1in

\noindent
{\bf Proof:} We start by noting that
\begin{eqnarray}
\kappa_{\text{Noether}} &\equiv& \Phi\, a_\mu \hat R^\mu
= \Phi\, v_{\mu ; \nu} v^\nu \hat R^\mu \nonumber \\
&=& \Phi_{ ; \nu} v_\mu v^\nu  \hat R^\mu +
 \Phi\, v_{\mu ; \nu} v^\nu \hat R^\mu -\Phi_{;\mu} \hat R^\mu \nonumber \\
&& -  \Phi_{ ; \mu} v_\nu v^\nu  \hat R^\mu
-  \Phi\, v_{\nu ; \mu} v^\nu  \hat R^\mu  \nonumber \\
&=& (\Phi\, v_\mu)_{ ; \nu} v^\nu  \hat R^\mu
 -\Phi_{;\mu} \hat R^\mu
 - (\Phi\, v_\nu)_{ ; \mu} v^\nu  \hat R^\mu \nonumber \\
&=& 2 (\Phi\, v_{[\mu})_{; \nu]} v^\nu  \hat R^\mu -\Phi_{;\mu} \hat R^\mu,
\label{lemma1.1}
\end{eqnarray}
where to obtain the second line we use the facts that
$v_\mu \hat R^\mu=0$, $v_\nu v^\nu=-1$ and hence $v_{\nu;\mu} v^\nu=0$.
Noting that \begin{equation}
\xi^\mu \equiv \frac{d x^\mu(\tau)}{dt}
= \frac{d\tau}{dt}\frac{dx^\mu(\tau)}{d\tau}
=\Phi\, v^\mu,
\end{equation}
we easily see that $\xi^\mu \xi_\mu =-\Phi^2$. It follows that
\begin{eqnarray}
 (\xi^\mu \xi_\mu)_{;\nu} \hat R^\nu &=&-(\Phi^2)_{;\mu} \hat R^\mu 
\nonumber \\
\Rightarrow~~~
2\,\xi^\mu_{~\,;\nu} \xi_\mu \hat R^\nu &=&-2\,\Phi_{;\mu} \Phi \hat R^\mu 
\nonumber \\
\Rightarrow~~~~~
\xi^\mu_{~\,;\nu} v_\mu \hat R^\nu &=&-\Phi_{;\mu} \hat R^\mu .
\label{extra}
\end{eqnarray}

Next, recalling that the Lie derivative of a vector is given by
$\pounds_\xi( \hat R^\mu )
\equiv \hat R^\mu_{~\,;\nu}\, \xi^\nu -\xi^\mu_{~\,;\nu} \hat R^\nu$, 
Eq.~(\ref{extra}) becomes
\begin{eqnarray}
\Phi_{;\mu} \hat R^\mu 
&=& -\xi^\mu_{~\,;\nu} v_\mu \hat R^\nu 
= \pounds_\xi (\hat R^\mu ) \,v_\mu
-\hat R^\mu_{~\,;\nu}\, \xi^\nu v_\mu \nonumber \\
&=& \pounds_\xi (\hat R^\mu ) \,v_\mu
-\big(\hat R^\mu v_\mu\big)_{;\nu} \xi^\nu 
+\hat R^\mu v_{\mu;\nu} \,\xi^\nu \nonumber \\
&=& \pounds_\xi (\hat R^\mu )\, v_\mu
+v_{\mu;\nu} \Phi \,v^\nu \hat R^\mu \nonumber \\
&=& \Phi \,a_\mu \hat R^\mu +\pounds_\xi (\hat R^\mu )\, v_\mu,
\label{extra2}
\end{eqnarray}
where we used $\hat R^\mu v_\mu=0$ and $\xi^\nu =\Phi \,v^\nu$ to obtain
line three. To remove the Lie derivative term one
might expect to have to apply a Killing condition. However, this is
unnecessary since we may always describe $\hat R^\mu$ as the
tangent vector to some spacelike path $x^\mu(s)$ parameterized by proper
length $s$, so
\begin{equation}
\hat R^\mu = \frac{dx^\mu(s)}{ds}.
\end{equation}
In particular, recalling that by construction $\xi^\mu =dx^\mu /dt$,
we find \cite{Poisson2004}
\begin{eqnarray}
\pounds_\xi (\hat R^\mu )
&\equiv& \hat R^\mu_{~\,;\nu}\, \xi^\nu -\xi^\mu_{~\,;\nu} \hat R^\nu
=\hat R^\mu_{~\,,\nu}\, \xi^\nu -\xi^\mu_{~\,,\nu} \hat R^\nu \nonumber \\
&=& \frac{\partial}{\partial x^\nu} \Bigl(\frac{dx^\mu}{ds}\Bigr)
\frac{dx^\nu}{dt}
-\frac{\partial}{\partial x^\nu} \Bigl(\frac{dx^\mu}{dt}\Bigr)
\frac{dx^\nu}{ds}
\nonumber \\
&=& \frac{ d^2 x^\mu}{dt ds} - \frac{ d^2 x^\mu}{ds dt} \equiv 0,
\end{eqnarray}
where in the final step, the fact that the coordinates $x^\mu$ are scalar
functions allows us to exchange the order of differentiation.
Thus Eq.~(\ref{extra2}) reduces to 
\begin{equation}
\Phi_{;\mu} \hat R^\mu= \Phi
\,a_\mu \hat R^\mu = \kappa_{\text{Noether}}
\label{10}
\end{equation}
and hence Eq.~(\ref{lemma1.1}) becomes
\begin{eqnarray}
\kappa_{\text{Noether}} \,dA
&=& (\Phi\, v_{[\mu})_{; \nu]} v^\nu  \hat R^\mu dA \nonumber \\
&=& \xi_{[\mu; \nu]} v^{[\nu}  \hat R^{\mu]} dA \nonumber \\
&=& \xi_{[\mu; \nu]} \hat T^{[\nu}  \hat N^{\mu]} dA 
=  \xi^{[\mu; \nu]}d\Sigma_{\mu\nu} .
\end{eqnarray}
Here to obtain the second line we rely on the identification of $\xi^\mu
= \Phi\, v^\mu$ and the anti-symmetry already present in the indices
$(\mu, \nu)$. To obtain the third line note that
the 4-velocity $v^\mu$ is orthogonal to the tangent space of ${\cal
S}$. Thus, the pair of unit vectors $\{v^\mu,\hat R^\mu\}$
span the same 2-dimensional tangent space as the pair of unit vectors
$\{\hat T^\mu,\hat N^\mu\}$. Finally, for the 2-surface
${\cal S}$ in the hypersurface $\Sigma_t$, we have $d\Sigma_{\mu\nu}=\hat
T_{[\nu}  \hat N_{\mu]} dA$. \qed

\vskip 0.1in

We have now derived the naturally defined surface gravity for a family
of observers and shown that it is the Noether charge,
$\kappa_{\text{Noether}}$. The construction of the Noether charge from
(\ref{kappaN}) is not limited to stationary spacetimes, nor does this
Theorem require the existence of a Killing vector. The generality of
this result and its physical interpretation both derive from the
scenario we considered: on the one hand, a general timelike
world tube, and on the other, an observer on the 2-surface of that world
tube, who therefore feels a physical surface gravity. Together, the
generality and physical grounding of our result provide the in-principle
possibility of direct observational access to a gravitational Noether
charge in dynamical spacetimes. 
% NETTA NOT HAPPY WITH ABOVE SENTENCE. REVISIT

By choosing differing world tubes one can address different physical
questions. To conclude this section, we consider the surface gravity of
spacetime horizons, where the choice of world tube is unambiguous.
Timelike horizons are a special case of the timelike world tubes
considered above, and may be treated in the same way. To derive the
surface gravity, one need only choose the world tube ${\cal W}$ as the
horizon world tube and apply the expression for
$\kappa_{\text{Noether}}$. The case of null horizons is considered next
(spacelike horizons are not considered in this paper).

\paragraph*{Surface gravity for null horizons:}

For world tubes of null horizons, one may take the local surface gravity
to be the limit as the congruence of observers approaches the null world
tube on the hypersurface of interest. In particular, we now show that in
the stationary case, this surface gravity reduces to the standard
result.

Consider a non-degenerate Killing
horizon with Killing vector
\begin{equation}
K^\mu = (\partial_t)^\mu + \Omega_\text{H}\, (\partial_\phi)^\mu,
\end{equation}
where $t$ is the coordinate time at spatial infinity, $\phi$ is the
azimuthal angular coordinate, and the constant $\Omega_\text{H}$ is the
angular velocity of the horizon for a Kerr black hole. According to the
conventional definition, the surface gravity, $\kappa_\text{Killing}$,
for such a Killing horizon satisfies\cite{Poisson2004}
\begin{equation}
{K^\mu}_{;\nu} K^\nu = \kappa_\text{Killing} \, K^\mu ,
\label{kappaK}
\end{equation}
on the horizon.

\vskip 0.1in
\noindent
{\bf Lemma 1:} The Noether surface gravity, $\kappa_\text{Noether}$,
reduces to the standard result, $\kappa_\text{Killing}$, for
non-degenerate Killing horizons.

\vskip 0.1in
The detailed proof is provided in the Appendix 1.

\section*{Gravitating mass as an integrated Noether charge}

Consider an observer, Albert, on a 2-surface ${\cal S}$ of a general
timelike world tube. As already mentioned, $a_\mu \hat R^\mu$ is the
component of the acceleration normal to ${\cal S}$ in Albert's own rest
frame. It can be thought of as the force-per-unit-mass required to keep
Albert on his world tube. Viewed by the reference observer, Emmy, the
Noether charge, $\Phi\, a_\mu \hat R^\mu$, is the force-per-unit-mass
that Emmy must exert on Albert to keep him on his world tube.
Thus, Emmy's Noether charge provides a common reference for measuring
the normal component of force for any accelerating observer on the world
tube. Integrating these forces over ${\cal S}$ yields the flux of force
normal to ${\cal S}$, which we interpret as the `gravitating mass'
responsible for the acceleration felt by a congruence of such observers. 

\vskip 0.1in

\noindent
{\bf Corollary 1:} The `gravitating mass' is defined by
\begin{equation}
  M_{\text{Grav}} \equiv \frac{1}{4\pi} \int_{\cal S}
 \Phi \,a_\mu \hat R^\mu dA
=\frac{1}{4\pi} \int_{\cal S}
 \kappa_{\text{Noether}}\, dA \; .
 \label{thm}
\end{equation}

\vskip 0.1in
For example, for `stationary' observers anywhere outside of the
horizon of a Schwarzschild black hole of mass $M$, Eq.~(\ref{thm})
reduces to $M_{\text{Grav}}=M$.

Quasi-local energies take the form of integrals over a 2-surface, 
e.g., the Brown-York mass (see Appendix 2). While $M_{\text{Grav}}$
therefore appears to be a quasi-local energy, Eq.~(\ref{Stokes}) allows 
it to be recast as a volume integral of
\begin{equation}
\rho_{\text{Grav}}\equiv \frac{1}{4\pi}\,J^\mu\hat T_\mu,
\label{ed1}
\end{equation}
i.e., the time-component of our locally conserved current $J^\mu$. We
therefore dub $\rho_{\text{Grav}}$ the `gravitating energy density.' To
simplify Eq.~(\ref{ed1}), it is convenient to first restate Eq.~(\ref{10})
as a general result:

\vskip 0.1in

\noindent
{\bf Corollary 2:} For our family of observers
\begin{equation}
\kappa_{\text{Noether}}= \Phi_{;\mu} \hat R^\mu.
% = \Phi \,a_\mu \hat R^\mu = \kappa_{\text{Noether}}
\label{redshift}
\end{equation}

\vskip 0.1in

The gravitating energy density reduces to a particularly simple form
when we consider world tubes $\{{\cal W}\}$, each of whose observers are
at `rest' with respect to Emmy's preferred foliation of the spacetime
$\Sigma_t$. In curved spacetime, we might consider an observer,
Leonhard, to be at rest when the nearby events he classifies as
simultaneous are likewise classified by Emmy (and so lie on $\Sigma_t$).
Such observers are called Eulerian observers and their 4-velocities must
be normal to the foliation,\cite{Gour2012} so $v^\mu=\hat T^\mu$. In
Leonhard's rest frame, the vector orthogonal to the 2-surface ${\cal S}$
will therefore also be tangent to $\Sigma_t$, so that $\hat R^\mu=\hat
N^\mu$. Then, Corollary~2 yields
\begin{equation}
\kappa_{\text{Noether}} =\hat N^\mu \nabla_\mu \Phi_{\text{Euler}}
= \hat N^a D_a\Phi_{\text{Euler}},
\end{equation}
where we explicitly label the red-shift factors of our Eulerian
observers as $\Phi_{\text{Euler}}$, and $D_a$ and $\hat N^a$ denote the
covariant derivative $\nabla_\mu$ and hypersurface-tangent 4-vector
$\hat N^\mu$, respectively, projected onto the 3-dimensional manifold of
the hypersurface $\Sigma_t$ with metric $h_{ab}$. Stokes' theorem on
this 3-manifold then yields
\begin{equation}
\rho_{\text{Grav}} = \frac{1}{4\pi} D^a D_a \Phi_{\text{Euler}},
\label{Poisson}
\end{equation}
where $D^a\equiv h^{ab}D_b$ and $h^{ab}$ is just the inverse matrix of
$h_{ab}$. We see, therefore, that {\it given access to the red-shift
factors alone of distributed families of Eulerian observers,
we could determine the gravitating energy density\/} from
Eq.~(\ref{Poisson}). Curiously, this relationship is reminiscent of
Poisson's law for Newtonian gravitation on a curved 3-geometry, where
the role of the gravitational potential is played by the scalar field of
red-shift factors, $\Phi_{\text{Euler}}$, for families of Eulerian
observers. Note that Eq.~(\ref{Poisson}) holds in arbitrary dynamical
spacetimes and is an exact relation rather than a slow-motion
weak-field approximation of general relativity.\cite{Carroll}

Note that the gravitating mass depends on the choice of the family of
observers. In fact such behavior is required by consistency with the
equivalence principle for such an observable quantity. For example, for
families of geodesic observers the acceleration is zero and hence for
them the gravitating mass vanishes.

One might worry that although fixing the observers' world tube to be
orthogonal to the foliating hypersurfaces we have merely shifted the
arbitrariness into the choice of how we foliate the spacetime. In
practice, this choice of foliation is made naturally by the astronomer
herself --- by the reference observer Emmy. The Noether charge
formulation allows Emmy, in-principle, to directly access the detailed
distribution of gravitating energy density, whenever she has access to
the red-shift factors of non-geodesic Eulerian observers, whose
description of simultaneity agrees with her own. Unfortunately, as no
such observers likely exist. To make use of the above formalism, Emmy
must rely on information from geodesic observers as surrogates to
extract the data needed. We now describe an approach Emmy might use to
do this.

% \onecolumngrid
% \begin{center}
\begin{figure}[ht]
\centering
\vskip -0.06truein
\includegraphics[width=0.48\textwidth]{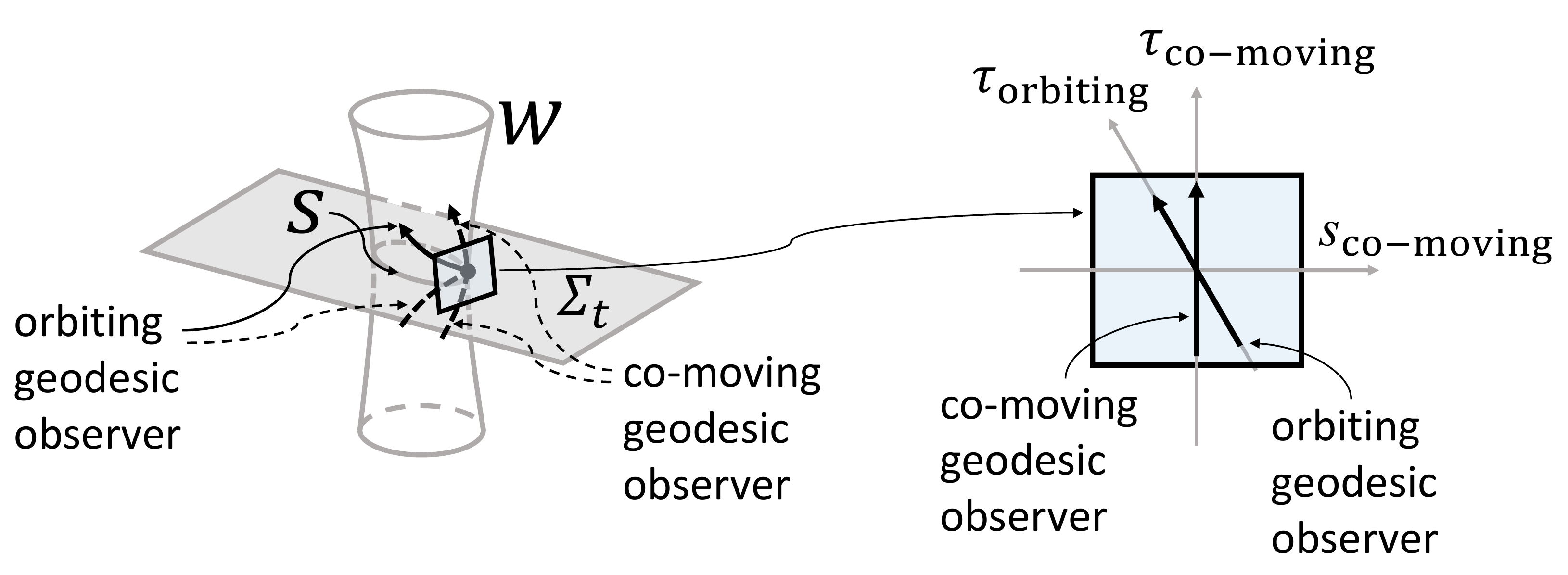}
\vskip -0.1truein

\caption{The hypersurface $\Sigma_t$, world tube ${\cal W}$ and their
intersection 2-surface ${\cal S}$ are all in faint gray. The
non-geodesic observer (not shown) follows the world tube. The co-moving
geodesic observer is on a parabolic-like arc kissing the world tube (and
tangent to the non-geodesic observer's trajectory) at one spacetime
point. In this illustration, the orbiting geodesic observer is hugging
the outside of the world tube as it spirals around the central region.
This orbital geodesic crosses the other two trajectories at their
spacetime point of intersection.}

\vskip -0.1truein
\label{Observers}
\end{figure}
% \end{center}
% \twocolumngrid

\section*{Noether charge astronomy}

At first sight Emmy's task seems hopeless. As noted above, geodesic
observers by their very nature are in free fall and hence feel no
acceleration and consequently have vanishing surface gravity.
Nevertheless, according to Corollary~2, it is sufficient for Emmy's
purposes to determine what would be the red-shift factors of a set of
fictitious non-geodesic Eulerian observers. To achieve this she may rely
on the following:

\vskip 0.1in

\noindent
{\bf Observation:} The red-shift factor is solely a function of the
observer's instantaneous 4-velocity $v^\mu$. In particular,
\begin{equation}
\Phi = \frac{1}{v^t}.
\end{equation}

% \vskip 0.1in

\noindent
{\bf Proof:}
Within a proper time $\delta \tau$ this
observer has moved by $\delta x^\mu =v^\mu \delta\tau$. The
$t$-component of this expression is $\delta t = v^t \delta \tau$,
from which the observation follows.
\qed

% \vskip 0.1in

Since the red-shift factors only depend on the instantaneous velocity,
Emmy may replace the non-geodesic observer with a co-moving geodesic
observer with instantaneous common 4-velocity (see
Fig.~\ref{Observers}). Next, she may relate the red-shift factor of this
co-moving observer with that of an orbiting observer crossing the same
spacetime point. She can achieve this by performing a Lorentz
transformation on the red-shift factor for the orbiting geodesic to
instantaneously (de)boost it to that of the co-moving observer (see
Fig.~\ref{Observers}). The former replacement does not change the
red-shift factor while the latter involves a Lorentz transformation
factor (see Appendix 1).

In this way, Emmy may extract the red-shift factor of a fictitious
non-geodesic observer from that of an orbiting observer at the same
spacetime point. The orbiting observer is presumed to be emitting light
with a well-characterized spectrum, e.g., it might be a star. In
principle then every star could provide a point in a distribution of
red-shift factors, from which Emmy may determine the local
Noether-charge surface gravity for her fictitious observers, or
equivalently the gravitating energy density, Eq.~(\ref{Poisson}), or net
gravitating mass, Eq.~(\ref{thm}), across the population of stars
consistent with how she sees the Universe.

\section*{Discussion}

We derive a naturally defined surface gravity using a congruence of
timelike observers following a world tube. We show that this surface
gravity is a Noether charge. The associated Noether current is conserved
in arbitrarily dynamical spacetimes and does not require the Einstein
field equations to hold true (see Appendix 1). Thus, this Noether charge
has a fundamental character which likely survives even in quantum
gravity where quantum fluctuations are expected to make the classical
field equations only a lowest-order approximation.

This Noether-charge surface gravity may be extended to null world tubes
by considering the limit as the timelike world tubes locally approach
such null hypersurfaces, in which case the surface gravity reduces to
the standard result on Killing horizons. Its fundamental nature makes this
Noether-charge formulation a natural candidate for the surface gravity of 
{\it dynamical\/} horizons (whether timelike or null). There are a number 
of alternative candidates for this quantity (see Appendix 2). However, 
there is as yet no consensus even for the simplest dynamical horizons 
of spherically-symmetric black holes.\cite{Nielsen2008}

We show that the surface gravity is in-principle observable, in this way
addressing the long standing challenge to identify a generally
observable gravitational Noether charge.\cite{Fletcher1960} We associate
this surface gravity with the gravitating mass (its integrated form). We
have shown that for asymptotically stationary observers (with $\xi^\mu
=\partial_t$), the gravitating mass reduces to the Komar energy.
Thus, for the first time we may provide a physically measurable
interpretation to the Komar energy even in dynamical spacetimes. The
existence of such an interpretation has been long
questioned.\cite{Bak1994} We also note that the integrated Noether
charge has been previously conjectured to be the entropy of dynamical
spacetime horizons even for generalized theories of
gravity,\cite{Wald1993} rather than as the energy found here. However,
this conjecture is based on scaling away the surface gravity from the
Noether charge (see Appendix 2).

Finally, we have proposed a scheme by which this Noether charge
formalism could be used to probe the detailed distribution of
gravitating mass within galaxies, galaxy clusters etc., from the
red-shift factors of well-characterized sources following orbits in such
systems. Such a new and independent probe may lead to novel insights
into the nature of dark matter in our universe.

\section*{APPENDIX}

This supplementary information is separated into two appendices. In
Appendix 1, we give detailed proofs for claims in the main manuscript.
In Appendix 2, we look at previous work with connections to our results.

\section*{APPENDIX 1: DETAILED PROOFS}

\section*{Hamiltonian formulation of Noether's theorem}

In the classical domain, Noether's theorem shows that if evolution
generated (in the sense of the Poisson bracket formalism) by the system
Hamiltonian leaves some quantity invariant then the evolution generated
by that quantity will leave the system Hamiltonian invariant and vice
versa.\cite{Noether1918Ap} In order make this result accessible we
repeat Baez's proof here.\cite{Baez2018Ap}

\vskip 0.1in
\noindent
{\bf Proof:} Poisson brackets allow us to construct the one-parameter
evolution generated by a `Hamiltonian' $H$ on some quantity $a$ via
\begin{equation}
\frac{da(t)}{dt} = \left\{ H, a(t) \right\},
\end{equation}

Suppose some quantity $a$ is invariant under this evolution, then
\begin{eqnarray}
&&a(t)=a(t=0)\equiv a \nonumber \\
&\Leftrightarrow& \frac{da(t)}{dt} = 0\nonumber \\
&\Leftrightarrow& \left\{ H, a(t) \right\} = 0\nonumber \\
&\Leftrightarrow& \left\{ H, a \right\} = 0\nonumber \\
&\Leftrightarrow& \left\{ a, H \right\} = 0.
\end{eqnarray}
Therefore the one-parameter action generated by $a$ on the system
Hamiltonian also vanishes, leaving $H$ invariant under this action. \qed

Such invariant quantities are called Noether charges. We note in passing
that Noether's theorem generalizes without difficulty to standard
quantum mechanics, but fails to hold for real or quaternionic versions
of quantum mechanics.\cite{Baez2018Ap}

\section*{Conservation of the Komar current density}

In 1959, within the context of general relativity, Komar showed for
an arbitrary vector field $\xi^\mu$, that the
4-current $J^\mu(\xi)$ is locally conserved,\cite{Komar1959Ap} where
\begin{equation}
J^\mu(\xi ) \equiv {\xi^{[\nu;\mu]}}_{;\nu} \;. 
\end{equation}
Here, we will show that 
\vskip 0.1in

\noindent
{\bf Lemma 0:} ${J^\mu(\xi)}_{;\mu}=0$ for any theory
with a torsion-free metric-compatible connection.

\vskip 0.1in

\noindent
{\bf Proof:} To simplify  $J^\mu(\xi)$, we first introduce the
general covariant derivative formula with a non-zero torsion tensor,
defined as $T^\alpha_{\mu\nu}\equiv
\Gamma^\alpha_{\mu\nu} - \Gamma^\alpha_{\nu\mu}$.
Indeed, the torsion tensor comes from the presumed non-commutativity
of the covariant derivative of a scalar function
\begin{eqnarray}
f_{;\mu \nu} - f_{;\nu \mu} &=& \nabla_\nu \nabla_\mu f
- \nabla_\mu \nabla_\nu f \nonumber \\
&=& f_{,\mu\nu} - \Gamma^\lambda_{\nu \mu} f_{,\lambda}
- (f_{,\nu\mu} - \Gamma^\lambda_{\mu \nu} f_{,\lambda} )\nonumber \\
&=& f_{,\mu\nu} - f_{,\nu\mu} - \Gamma^\lambda_{\nu \mu} f_{,\lambda}
+ \Gamma^\lambda_{\mu \nu} f_{,\lambda} \nonumber \\
&=&  - \Gamma^\lambda_{\nu \mu} f_{,\lambda}
+ \Gamma^\lambda_{\mu \nu} f_{,\lambda} \nonumber \\
&=& T^\lambda_{\mu\nu} f_{,\lambda} \,,
\label{defTorsion}
\end{eqnarray}
where we have assumed ordinary derivatives are commutative to obtain
the forth line.
Based on this definition of torsion tensor, it is known
that\cite{carroll2004Ap}
\begin{eqnarray}
2{X^{\mu_1 \cdots \mu_k}}_{\nu_1 \cdots \nu_l ;[\alpha \beta]}
&=& T^\lambda_{\alpha \beta}{X^{\mu_1 \cdots \mu_k}}_{\nu_1 \cdots \nu_l
; \lambda} \nonumber \\ 
  &&+ {R^\lambda}_{\nu_1 \alpha \beta}
{X^{\mu_1 \cdots \mu_k}}_{\lambda \cdots \nu_l} + \cdots \nonumber \\ 
  &&- {R^{\mu_1}}_{\lambda \alpha \beta}
{X^{\lambda \cdots \mu_k}}_{\nu_1 \cdots \nu_l} -\cdots
\label{Coderi}
\end{eqnarray}
From Eq.~(\ref{Coderi}), we can see that the last two indices of
the Riemann tensor are anti-symmetric.

Thus, the covariant divergence of $J^\mu(\xi)$ may be simplified
as follows
\begin{eqnarray}
  2{J^\mu(\xi)}_{;\mu} &=& 2{{\xi^{[\nu;\mu]}}}_{;\nu\mu} 
  \nonumber \\ 
  &=& 
(\xi ^{\nu;\mu})_{;\nu\mu}  - (\xi^{\mu;\nu})_{;\nu\mu} \nonumber \\ 
  &=& (\xi ^{\nu;\mu})_{;\nu\mu}  -  (\xi^{\nu;\mu})_{;\mu\nu} 
  \nonumber \\ 
  &=&  T^\alpha_{\nu\mu} {\xi^{\nu ; \mu}}_{; \alpha}
-  {R^\nu}_{ \alpha \nu \mu} \xi^{\alpha ; \mu}
- {R^\mu}_{ \alpha \nu \mu} \xi^{\nu ; \alpha} \nonumber \\
    &=&  T^\alpha_{\nu\mu} {\xi^{\nu ; \mu}}_{; \alpha}
- {R^\nu}_{ \alpha \nu \mu} \xi^{\alpha ; \mu}
+ {R^\mu}_{ \alpha \mu \nu} \xi^{\nu ; \alpha}  \nonumber \\ 
  &=& T^\alpha_{\nu\mu} {\xi^{\nu ; \mu}}_{; \alpha}
+  R_{ \alpha \nu} \xi^{\nu ;\alpha}
-  R_{ \alpha \mu}  \xi^{\alpha ; \mu} \nonumber \\ 
  &=& T^\alpha_{\nu\mu} {\xi^{\nu ; \mu}}_{; \alpha}
+  ( R_{ \alpha \nu} -R_{\nu \alpha} )\xi^{\nu ;\alpha} ,
  \label{con00}
\end{eqnarray}
where we have used Eq.~(\ref{Coderi}) to obtain line four, and we have
used the property that the last two indices of
${R^\mu}_{\nu \alpha \beta}$ are anti-symmetric to obtain line five. 

Since $\xi^\mu$ is arbitrary and the first term and the second term in
the last line of Eq.~(\ref{con00}) have different orders of derivative
of $\xi^\mu$, to ensure ${J^\mu(\xi)}_{;\mu}$ vanishes, we expect to
independently require that $T^\alpha_{\nu\mu}=0$ and $R_{ \alpha \nu}
=R_{\nu \alpha}$. While the vanishing of $T^\alpha_{\nu\mu}$ means that
the geometry is torsion-free, it is not so
clear what $R_{ \alpha \nu} -R_{\nu \alpha}=0$ represents physically.
We shall analyze this now. 

We first generalize the Bianchi symmetry to a geometry with a non-vanishing
torsion tensor. \cite{Penrose1984Ap} To derive this relation, let us
first expand $2 f_{;[\mu \nu \alpha]}$ with the first two indices
kept anti-symmetric, yielding
\begin{eqnarray}
2 f_{;[\mu \nu \alpha]}
&=& \frac{2}{3} f_{;[\mu \nu] ;\alpha}
+ \frac{2}{3} f_{;[\alpha  \mu] ; \nu }
+ \frac{2}{3} f_{;[ \nu \alpha] ; \mu}   \nonumber \\ 
  &=& \frac{1}{3} (T^\lambda_{\mu \nu} f_{;\lambda})_{;\alpha}
+ \frac{1}{3} (T^\lambda_{\alpha \mu} f_{;\lambda})_{;\nu }
+ \frac{1}{3} (T^\lambda_{\nu\alpha} f_{;\lambda})_{; \mu}   \nonumber \\ 
  &=& \frac{1}{3} T^\lambda_{\mu \nu ;\alpha} f_{;\lambda}
+ \frac{1}{3} T^\lambda_{\alpha \mu ;\nu} f_{;\lambda}
+ \frac{1}{3} T^\lambda_{\nu \alpha ; \mu} f_{;\lambda}   \nonumber \\ 
  &&+ \frac{1}{3} T^\lambda_{\mu \nu} f_{;\lambda \alpha}
+ \frac{1}{3} T^\lambda_{\alpha \mu} f_{;\lambda \nu }
+ \frac{1}{3} T^\lambda_{\nu\alpha} f_{;\lambda \mu}  \nonumber \\ 
  &=& T^\lambda_{[\mu \nu ;\alpha]} f_{;\lambda}  \nonumber \\ 
  &&+ \frac{1}{3} T^\lambda_{\mu \nu} f_{;\lambda \alpha}
+ \frac{1}{3} T^\lambda_{\alpha \mu} f_{;\lambda \nu }
+ \frac{1}{3} T^\lambda_{\nu\alpha} f_{;\lambda \mu} ,
\label{fmna1}
\end{eqnarray}
where we have used Eq.~(\ref{defTorsion}) to obtain line two.
On the other hand, we may expand $2 f_{;[\mu \nu \alpha]}$ 
keeping the last two indices anti-symmetric to yield
\begin{eqnarray}
2 f_{;[\mu \nu \alpha]} &=& \frac{2}{3} f_{;\mu ; [\nu \alpha]}
+ \frac{2}{3} f_{;\alpha ; [\mu \nu] }
+ \frac{2}{3} f_{; \nu ; [\alpha \mu]}   \nonumber \\ 
  &=& \frac{1}{3} (T^\lambda_{\nu \alpha} f_{;\mu \lambda}
+ {R^\lambda}_{\mu \nu \alpha} f_{;\lambda} )
+ \frac{1}{3} (T^\lambda_{\mu \nu} f_{;\alpha \lambda } \nonumber \\ 
  &&+ {R^\lambda}_{\alpha \mu \nu } f_{;\lambda} )
+ \frac{1}{3} (T^\lambda_{\alpha \mu} f_{;\nu \lambda}
+ {R^\lambda}_{\nu \alpha \mu } f_{;\lambda} )  \nonumber \\ 
  &=& {R^\lambda}_{[\mu \nu \alpha]} f_{;\lambda} \nonumber \\ 
  &&+ \frac{1}{3} T^\lambda_{\nu \alpha} f_{;\mu \lambda}
+ \frac{1}{3} T^\lambda_{\mu \nu} f_{;\alpha \lambda }
+ \frac{1}{3} T^\lambda_{\alpha \mu} f_{;\nu \lambda}  ,
\label{fmna2}
\end{eqnarray}
where we have applied Eq.~(\ref{Coderi}) to obtain line two. 

Since Eqs.~(\ref{fmna1}) and~(\ref{fmna2}) are equal, we have 
\begin{eqnarray}
&&{R^\lambda}_{[\mu \nu \alpha]} f_{;\lambda}
- T^\lambda_{[\mu \nu ;\alpha]} f_{;\lambda}  \nonumber \\ 
  &=&\frac{1}{3} T^\lambda_{\mu \nu} f_{;\lambda \alpha}
+ \frac{1}{3} T^\lambda_{\alpha \mu} f_{;\lambda \nu }
+ \frac{1}{3} T^\lambda_{\nu\alpha} f_{;\lambda \mu}   \nonumber \\
  &&-( \frac{1}{3} T^\lambda_{\nu \alpha} f_{;\mu \lambda}
+ \frac{1}{3} T^\lambda_{\mu \nu} f_{;\alpha \lambda }
+ \frac{1}{3} T^\lambda_{\alpha \mu} f_{;\nu \lambda} ) \nonumber \\ 
  &=& \frac{1}{3} T^\lambda_{\mu \nu} T^\tau _{\lambda \alpha} f_{;\tau}
+ \frac{1}{3} T^\lambda_{\alpha \mu}  T^\tau _{\lambda \nu} f_{;\tau}
+ \frac{1}{3} T^\lambda_{\nu\alpha} T^\tau _{\lambda \mu} f_{;\tau}
 \nonumber \\ 
  &=& - T^\lambda_{[\mu \nu} T^\tau _{\alpha ] \lambda } f_{;\tau}
= - T^\tau_{[\mu \nu} T^\lambda _{\alpha ] \tau } f_{;\lambda} \,,
\end{eqnarray}
where we have applied Eq.~(\ref{defTorsion}) to obtain line three.
Since $f$ is an arbitrary scalar function, the Bianchi identity with
a non-vanishing torsion tensor may be written
\begin{equation}
{R^\lambda}_{[\mu \nu \alpha]} - T^\lambda_{[\mu \nu ; \alpha ]}
+ T^\tau _{[\mu \nu} T^\lambda _{\alpha ] \tau} = 0 .
\label{Bianchi}
\end{equation}
(Note that there is a minus sign difference before the second term
between our result and that of Ref.~[\onlinecite{Penrose1984Ap}].)

Expanding out the anti-symmetric relation in
${R^\lambda}_{[\mu \nu \alpha]}$, Eq.~(\ref{Bianchi}) becomes
\begin{eqnarray}
\frac{1}{3} ( {R^\lambda}_{\mu \nu \alpha}
+ {R^\lambda}_{ \nu \alpha \mu }
+ {R^\lambda}_{\alpha \mu \nu }) 
= T^\lambda_{[\mu \nu ; \alpha ]}
- T^\tau _{[\mu \nu} T^\lambda _{\alpha ] \tau} .
\label{Bianchi2}
\end{eqnarray}
Contracting the indices $\lambda$ and $\nu$ in
Eq.~(\ref{Bianchi2}) then yields
\begin{eqnarray}
{R^\lambda}_{\mu \lambda \alpha}
+ {R^\lambda}_{ \lambda \alpha \mu }
+ {R^\lambda}_{\alpha \mu \lambda } 
&=& 3 T^\lambda_{[\mu \lambda ; \alpha ]}
- 3 T^\tau _{[\mu \lambda} T^\lambda _{\alpha ] \tau}  \nonumber \\
\Rightarrow   {R^\lambda}_{\mu \lambda \alpha}
+ {R^\lambda}_{ \lambda \alpha \mu }
- {R^\lambda}_{\alpha \lambda \mu} 
&=& 3 T^\lambda_{[\mu \lambda ; \alpha ]}
- 3 T^\tau _{[\mu \lambda} T^\lambda _{\alpha ] \tau}  \nonumber \\
\Rightarrow ~~~~~~  R_{\mu \alpha}
+ {R^\lambda}_{ \lambda \alpha \mu }
- R_{\alpha \mu} 
&=& 3 T^\lambda_{[\mu \lambda ; \alpha ]}
- 3 T^\tau _{[\mu \lambda} T^\lambda _{\alpha ] \tau}   \nonumber \\
\Rightarrow ~~~~~~  R_{\mu \alpha} - R_{\alpha \mu}
+ {R^\lambda}_{ \lambda \alpha \mu } 
&=& 3 T^\lambda_{[\mu \lambda ; \alpha ]}
- 3 T^\tau _{[\mu \lambda} T^\lambda _{\alpha ] \tau} . \nonumber \\
\label{Bianchi3}
\end{eqnarray}
Although ${R^\lambda}_{\lambda \alpha \mu}=0$ in general relativity
because the first two indices are anti-symmetric, this is not
necessary to be true for a generic geometry.

Inserting Eq.~(\ref{Bianchi3}) into Eq.~(\ref{con00}) yields
\begin{eqnarray}
  2{J^\mu(\xi)}_{;\mu} &=&  T^\alpha_{\nu\mu} {\xi^{\nu ; \mu}}_{; \alpha}
+ 3 ( T^\lambda_{[\alpha \lambda ; \nu ]}
- T^\tau _{[\alpha \lambda} T^\lambda _{\nu ] \tau} )\xi^{\nu ;\alpha}
\nonumber \\ 
  && -  {R^\lambda}_{\lambda \nu \alpha} \xi^{\nu ;\alpha}  .
  \label{con1}
\end{eqnarray}
To understand the geometric behavior of ${R^\lambda}_{\lambda \nu \alpha}$,
we now consider 
$2 g_{\mu\nu ;[\alpha \beta]}$
\begin{eqnarray}
  2 g_{\mu\nu ;[\alpha \beta]}
&=& T^\lambda_{\alpha \beta} \, g_{\mu\nu ;\lambda}
+ {R^\lambda}_{\mu\alpha \beta} g_{\lambda \nu} 
+ {R^\lambda}_{\nu\alpha \beta} g_{\mu\lambda}  \nonumber \\ 
  &=& T^\lambda_{\alpha \beta} \, g_{\mu\nu ;\lambda}
+ R_{\nu\mu\alpha \beta}  + R_{\mu\nu\alpha \beta}  .
  \label{metriccamp}
\end{eqnarray}
Contracting Eq.~(\ref{metriccamp}) with $g^{\mu\nu}$ then yields
\begin{eqnarray}
  2 g^{\mu\nu} g_{\mu\nu ;[\alpha \beta]}
&=& T^\lambda_{\alpha \beta} \, g^{\mu\nu} g_{\mu\nu ;\lambda}
+ 2 {R^\mu}_{\mu\alpha \beta}  .
  \label{metriccamp2}
\end{eqnarray}
Finally, inserting Eq.~(\ref{metriccamp2}) into Eq.~(\ref{con1}) gives
\begin{eqnarray}
  {J^\mu(\xi)}_{;\mu} \!\!
&=& \!\! \frac{1}{2} T^\alpha_{\nu\mu} {\xi^{\nu ; \mu}}_{; \alpha}
+ \frac{3}{2} ( T^\lambda_{[\alpha \lambda ; \nu ]}
- T^\tau _{[\alpha \lambda} T^\lambda _{\nu ] \tau} )\xi^{\nu ;\alpha}
\nonumber \\ 
  \!\! && \!\!\! +  \frac{1}{4} T^\lambda_{\nu \alpha}
\, g^{\mu\nu} g_{\mu\nu ;\lambda} \xi^{\nu ;\alpha}
\!-\! \frac{1}{2} g^{\mu\nu} g_{\mu\nu ;[\nu\alpha]} \xi^{\nu ;\alpha}  .
\end{eqnarray}

Since $\xi^\mu$ is an arbitrary vector and these terms are in different
orders of derivatives of $\xi^\mu$ or $T^\alpha_{\nu\mu}$, they
are independent of each other. Therefore, the most natural and
straightforward conditions to ensure that
\begin{eqnarray}
  {J^\mu(\xi)}_{;\mu} =0 ,
  \label{con0}
\end{eqnarray} 
are to require the torsion tensor vanish and the metric is compatible
with the covariant derivative (or connection of the covariant derivative). 

This completes the proof of Lemma 0.
\qed

\vskip 0.1in

The local conservation of $J^\mu(\xi)$ relies on the geometry of the
spacetime manifold; nothing else. It therefore allows us to construct a
family (one for each choice of vector field $\xi^\mu$) of locally
conserved quantities for any theory of gravity obeying these geometric
requirements. Furthermore, these conservation laws are preserved by any
Hamiltonian which respects the assumptions of being torsion free and
metric compatible. Therefore by Noether's theorem, we have an infinite
family of conserved currents and Noether charges for any 
torsion-free metric-compatible theory of gravity.

% \vskip 0.5in

\section*{Lagrangian formulation of Noether's theorem}

Consider a Lagrangian ${\cal L}(\phi,\phi_{,\mu})$ depending on the
field $\phi$ and its first-order derivative, then the
variation of ${\cal L}$ with respect to $\phi \rightarrow \phi+ \delta
\phi$ may be written
\begin{eqnarray}
\delta {\cal L} &=& \frac{\partial {\cal L}}{\partial \phi}
\delta \phi + \frac{\partial {\cal L}}{\partial (\phi_{,\mu})}
\delta (\phi_{,\mu})  \nonumber \\ 
  &=& \frac{\partial {\cal L}}{\partial \phi} \delta \phi
+ \Bigl( \frac{\partial {\cal L}}{\partial (\phi_{,\mu})}
\delta \phi \Bigr)_{\!,\mu}
\!- \Bigl( \frac{\partial {\cal L}}{\partial (\phi_{,\mu})} \Bigr)_{\!,\mu}
\delta \phi  \nonumber \\ 
  &=& \Bigl( \frac{\partial {\cal L}}{\partial (\phi_{,\mu})}
\delta \phi \Bigr)_{\!,\mu}\! + \Bigl( \frac{\partial {\cal L}}{\partial \phi}
 - \Bigl( \frac{\partial {\cal L}}
{\partial (\phi_{,\mu})} \Bigr)_{\!,\mu} \Bigr) \delta \phi .
  \label{Lagran1}
\end{eqnarray}

Here the second term vanishes, `on-shell,' due to the equations of
motion for the system. The first term is usually neglected
because $\delta \phi$ vanishes on the boundary. However, the expression
itself is generally non-zero. In fact Noether's theorem relies on this
term to define a locally conserved quantity when some symmetries exist.
In general, let us suppose that these symmetries are generated by
variations in some generalized coordinates $\theta^A$. Rewriting
the variation of ${\cal L}$ in terms of these coordinates we find
\begin{eqnarray}
\delta {\cal L} &=&  \frac{\partial {\cal L}}{\partial \theta^A}
\delta \theta^A \nonumber \\ 
  &=&  \frac{\partial ( {\cal L}\, \delta \theta^A )}{\partial \theta^A}
- {\cal L} \frac{\partial (\delta \theta^A) }{\partial \theta^A}  
  \nonumber \\ 
  &=&  \frac{\partial ( {\cal L} \,\delta \theta^A )}
{\partial \theta^A} \nonumber \\ 
  &=& ( {\cal L} \,\delta \theta^A )_{,A} .
\label{Lagran2}
\end{eqnarray}

Since Eqs.~(\ref{Lagran1}) and~(\ref{Lagran2}) both describe the variation
of the Lagrangian we have
\begin{eqnarray}
\Bigl( \frac{\partial {\cal L}}{\partial (\phi_{,\mu})}
\delta \phi \Bigr)_{,\mu} = ( {\cal L}\, \delta \theta^A )_{,A} ,
\label{Lagran3}
\end{eqnarray}
where we have assumed that the system is on-shell and hence obeys the
equations of motion so that the second term in Eq.~(\ref{Lagran1}) vanishes. 

From Eq.~(\ref{Lagran3}) we can see that here are two main possibilities
for the Noether's theorem. The first scenario is when the Lagrangian is
invariant with respect to the generalized coordinate $\theta^A$, so that
the right-hand-side of Eq.~(\ref{Lagran3}) vanishes. In this case, the
Noether current is defined simply as
\begin{equation}
J^\mu=\frac{\partial {\cal L}}{\partial (\phi_{,\mu})} \delta \phi.
\label{NoetherC1}
\end{equation}
The Lagrangian of a complex scalar field describes such a scenario
which we shall illustrate below.

The other main scenario we consider is when the generalized coordinates
$\theta^A$ reduce to the spacetime coordinates
$x^\mu$. In this case, Eq.~(\ref{Lagran3}) reduces to the on-shell
relation
\begin{eqnarray}
0 &=& \Bigl( \frac{\partial {\cal L}}{\partial (\phi_{,\mu})}
\delta \phi \Bigr)_{,\mu} - ( {\cal L} \,\delta x^\mu )_{,\mu}  \nonumber \\ 
  &=& \Bigl( \frac{\partial {\cal L}}{\partial (\phi_{,\mu})} \delta \phi
- {\cal L} \,\delta x^\mu \Bigr)_{,\mu} \nonumber \\ 
  &=& { J^\mu }_{,\mu} .
\label{Lagran4}
\end{eqnarray}
Here, the Noether current $J^\mu$ equals
\begin{equation}
J^\mu=\frac{\partial {\cal L}}{\partial (\phi_{,\mu})} \delta \phi
- {\cal L} \,\delta x^\mu.
\label{NoetherC2}
\end{equation}
Note that the current defined in this way will always be locally conserved
on-shell due to Eq.~(\ref{Lagran4}).

Now, let us consider the first of these scenarios in the case of the Lagrangian of
a complex scalar field
\begin{eqnarray}
{\cal L} = {\phi^\ast}_{,\mu} \phi^{,\mu} - m^2 \phi^\ast \phi .
\end{eqnarray}
For the variation $\phi \rightarrow \phi'=e^{i\theta}\phi $ the
Lagrangian does not change so there is a symmetry with respect to the
`gauge' parameter $\theta$. The variation of the Lagrangian with
respect to this parameter vanishes:
\begin{equation}
\delta {\cal L} = \frac{\partial {\cal L}}{\partial \theta}
\,\delta \theta = 0.
\end{equation} 
At the same time, the variations of the scalar fields with respect
to this gauge change are
\begin{eqnarray}
\delta \phi &=&  \phi' - \phi = (e^{i\theta} - 1) \phi \nonumber \\ 
\delta \phi^\ast &=&   (e^{-i\theta} - 1) \phi^\ast .
\label{phi}
\end{eqnarray}
Thus, based on Eq.~(\ref{NoetherC1}), the Noether current is 
\begin{eqnarray}
J^\mu &=&  \phi^{,\mu} (e^{-i\theta} - 1) \phi^\ast
+ \phi^{\ast,\mu} (e^{i\theta} - 1) \phi  .
\label{curren}
\end{eqnarray}
In the limit of small $\theta$ we have
$e^{i\theta} - 1= i \theta + O(\theta^2)$ and
$e^{-i\theta} - 1= -i \theta + O(\theta^2)$. In this limit,
Eq.~(\ref{curren}) may be further simplified to
\begin{eqnarray}
J^\mu &=&  i\theta (\phi^{,\mu} \phi^\ast - \phi^{\ast,\mu} \phi)  .
\label{curren2}
\end{eqnarray}
Using the Klein-Gordon equation, it is easy to check that this current
$J^\mu$ is locally conserved, having ${J^\mu}_{,\mu}=0$. Thus, we
see that the current of Eq.~(\ref{curren2}) is conserved on-shell.

All the above analysis and our example of the complex scalar field seem
to imply that a Noether current must be generically an on-shell
conserved current because we require the equations of motion to remove
the extra terms in $\delta {\cal L}$, e.g., the second term in
Eqs.~(\ref{Lagran1}). % and (\ref{LagranH1}). 
However, we shall see
below that the Noether current in a spacetime may be off-shell (meaning
independent of the validity of the equations of motion) because the
vanishing of these extra terms will be shown instead to be purely
geometric. In this way, the conservation of the Noether current is found
{\it not\/} to be limited to solutions of the Einstein field equations
of general relativity. Indeed, this appeared to be the natural
implication of the Hamiltonian formulation of the Noether current
described above. We shall now show that this result appears to hold true
explicitly within the Lagrangian formulation of the Noether current for
general relativity.

\section*{Lagrangian formulation of Noether's theorem for general relativity}

After having introduced the Noether current based on the Lagrangian
formulation above, we now derive the Noether current for general
relativity in this formulation.\cite{Iyer1994Ap,Bak1994Ap} In this section
and henceforth, we directly assume the spacetime geometry is
torsion-free and the covariant derivative is metric compatible. The
Lagrangian for general relativity is taken to be given by the
Einstein-Hilbert gravitational action with density $R\sqrt{-g}$. The
variation of this Lagrangian with respect to the metric $g_{\mu\nu}$ may
be calculated \cite{bertschinger2002Ap}
\begin{widetext}
\begin{eqnarray}
 \delta (R \sqrt{-g}) = -\Bigl( (R_{\mu \nu}-\frac{1}{2}g_{\mu \nu}R )
h^{\mu \nu} - (g^{\mu\nu} \delta {\Gamma ^\alpha} _{\mu \nu }
-g^{\mu\alpha } \delta {\Gamma ^\lambda} _{\lambda  \mu })_{;\alpha }
\Bigr) \sqrt{-g } ,
\label{VarRg}
\end{eqnarray}
where $h_{\mu\nu}\equiv \delta g_{\mu\nu}
= - g_{\mu\sigma} g_{\nu \tau} \delta g^{\sigma\tau}$,
i.e., $\delta g^{\sigma\tau} = -h^{\sigma\tau}$.

Now, we show that $-(g^{\mu\nu} \delta {\Gamma ^\alpha} _{\mu \nu }
-g^{\mu\alpha } \delta {\Gamma ^\lambda}  _{\lambda  \nu })_{;\alpha }
=2\, {{h^{\mu}}_{[\mu ; \nu]}}^{;\nu}$,
a result quoted in Ref.~[\onlinecite{Bardeen1973Ap}], there without proof.

Since $\delta \Gamma_{\mu \nu}^\alpha
= \frac{1}{2} g^{\alpha \rho } ( h_{\mu \rho ;\nu }
+ h_{\nu \rho ;\mu }-h_{\mu \nu ;\rho } )$,\cite{palatini1919Ap}
\begin{eqnarray}
    -(g^{\mu\nu} \delta {\Gamma ^\alpha} _{\mu \nu }-g^{\mu\alpha }
\delta {\Gamma ^\lambda}  _{\lambda  \mu })_{;\alpha } 
    &=& \Bigl( g^{\mu\alpha } \frac{1}{2} g^{\lambda \rho }
(  h_{\mu \rho ;\lambda } + h_{\lambda \rho ;\mu }-h_{\mu \lambda ;\rho } )
-   g^{\mu\nu } \frac{1}{2} g^{\alpha \rho } ( h_{\mu \rho ;\nu }
+ h_{\nu \rho ;\mu }-h_{\mu \nu ;\rho } ) \Bigr)_{;\alpha } 
    \nonumber  \\
    &=& \frac{1}{2} \Bigl( g^{\mu\alpha }  {h^\rho}_{\rho ;\mu }
-   g^{\mu\nu } ({h^\alpha}_{\mu ;\nu } + {h^\alpha}_{\nu ;\mu }
- {h_{\mu \nu}}^{;\alpha} ) \Bigr)_{;\alpha } 
    \nonumber  \\
    &=& \frac{1}{2} \Bigl( {h_\rho}^{\rho ;\alpha }
- ({h^{\alpha \nu}}_{;\nu } + {h^{\alpha \mu}}_{;\mu }
- {h_\mu}^{\mu ;\alpha} ) \Bigr)_{;\alpha }
    \nonumber  \\
    &=& \frac{1}{2} ( 2 \,{h_\rho}^{\rho ;\alpha }
- 2\, {h^{\alpha \mu}}_{;\mu } )_{;\alpha }
    \nonumber  \\
    &=& 2\, {{h^{\mu}}_{[\mu ; \nu]}}^{;\nu} .
    \label{delg}
\end{eqnarray}
Inserting Eq.~(\ref{delg}) into Eq.~(\ref{VarRg}) then yields
\begin{eqnarray}
 \delta (R \sqrt{-g}) = -\Bigl( (R_{\mu \nu}-\frac{1}{2}g_{\mu \nu}R )
h^{\mu \nu} + 2\, {{h^{\mu}}_{[\mu ; \nu]}}^{;\nu} \Bigr) \sqrt{-g } .
\label{VarR1}
\end{eqnarray}

Now consider the coordinate change: $x^\mu \rightarrow x'^\mu =
x^\mu + \xi^\mu$, so $\xi^\mu = \delta x^\mu$. It is not difficult
to see that the variation
of the metric under such a coordinate change is nothing but the Lie
derivative of the metric along $\xi^\mu$, yielding
\begin{equation}
h_{\mu\nu} =\delta g_{\mu\nu} = {\mathfrak{L}}_\xi g_{\mu\nu}
= 2 \xi_{(\mu ; \nu)}.
\label{LieMetric}
\end{equation}
Inserting Eq.~(\ref{LieMetric}) into the first term on the right-hand-side
of Eq.~(\ref{VarR1}) yields
\begin{eqnarray}
 ( R_{\mu \nu}-\frac{1}{2}g_{\mu \nu}R )
h^{\mu \nu} &=& (R_{\mu \nu}-\frac{1}{2}g_{\mu \nu}R ) \,
2  \xi^{(\mu ; \nu)} \nonumber  \\
    &=& (R_{\mu \nu}-\frac{1}{2}g_{\mu \nu}R ) \, 2  \xi^{\mu ; \nu}
 \nonumber  \\
    &=& 2 [(R_{\mu \nu}-\frac{1}{2}g_{\mu \nu}R ) \xi^\mu ]^{; \nu}
- 2 (R_{\mu \nu}-\frac{1}{2}g_{\mu \nu}R )^{ ; \nu}  \xi^\mu  .
\label{VarR2}
\end{eqnarray}
Similarly, inserting Eq.~(\ref{LieMetric}) into the second term on the
right-hand-side of Eq.~(\ref{VarR1}) yields
\begin{eqnarray}
2\, {{h^{\mu}}_{[\mu ; \nu]}}^{;\nu} &=& {{h^{\mu}}_{\mu ; \nu}}^{;\nu}
- {{h^{\mu}}_{\nu ; \mu}}^{;\nu}   \nonumber  \\
    &=& 2 {{\xi^\mu}_{;\mu \nu}}^{;\nu} - {\xi_{\mu ; \nu}}^{;\mu\nu}
- {\xi_{\nu ; \mu}}^{;\mu\nu}  \nonumber  \\
    &=& 2 {{\xi^\mu}_{;\mu \nu}}^{;\nu} - 2 {\xi_{\mu ; \nu}}^{;\mu\nu}
+ {\xi_{\mu ; \nu}}^{;\mu\nu} - {\xi_{\nu ; \mu}}^{;\mu\nu} \nonumber  \\
    &=& 2 ({\xi^\mu}_{;\mu \nu} - {\xi^\mu}_{;\nu \mu})^{;\nu}
+ 2{\xi_{[\mu ; \nu]}}^{;\mu\nu}  \nonumber  \\
    &=& 2(- R_{\mu\nu} \xi^\mu )^{;\nu} + 2{\xi_{[\mu ; \nu]}}^{;\mu\nu}  ,
\label{VarR3}
\end{eqnarray}
where we have used Eq.~(\ref{Coderi}) in moving from the fourth line to
the fifth line.

Combining the results of Eqs.~(\ref{VarR2}) and (\ref{VarR3}) into
Eq.~(\ref{VarR1}) gives us
\begin{eqnarray}
 \delta (R \sqrt{-g}) &=& -\Bigl( 2 ((R_{\mu \nu}-\frac{1}{2}
g_{\mu \nu}R ) \xi^\mu )^{; \nu} - 2 (R_{\mu \nu}
-\frac{1}{2}g_{\mu \nu}R )^{ ; \nu}  \xi^\mu
+ 2(- R_{\mu\nu} \xi^\mu )^{;\nu} + 2{\xi_{[\mu ; \nu]}}^{;\mu\nu}  \Bigr)
\sqrt{-g } \nonumber  \\
    &=& -\Bigl( - ( g_{\mu \nu}R \xi^\mu )^{; \nu}
- 2 (R_{\mu \nu}-\frac{1}{2}g_{\mu \nu}R )^{ ; \nu}  \xi^\mu
 + 2 {\xi_{[\mu ; \nu]}}^{;\mu\nu}  \Bigr) \sqrt{-g }  \nonumber  \\
    &=& ( R \xi^\mu )_{; \mu} \sqrt{-g } 
+ 2 {\xi^{[\mu ; \nu]}}_{;\mu\nu} \sqrt{-g } 
+ 2 (R_{\mu \nu}-\frac{1}{2}g_{\mu \nu}R )^{ ; \nu} \sqrt{-g }
\xi^\mu \nonumber  \\
    &=& ( R \xi^\mu \sqrt{-g } )_{, \mu} 
 + 2 ( {\xi^{[\mu ; \nu]}}_{;\mu } \sqrt{-g } )_{,\nu}
+ 2 (R_{\mu \nu}-\frac{1}{2}g_{\mu \nu}R )^{ ; \nu} \sqrt{-g } \xi^\mu  ,
\label{VarR}
\end{eqnarray}
where we have used ${A^\mu}_{;\mu} \sqrt{-g }
= (A^\mu \sqrt{-g })_{,\mu} $ for the first two terms in the final
step.\cite{Poisson2004Ap}
% \newline
% DOESN'T THIS ALSO REQUIRE 
% ${A^{\mu\nu}}_{;\mu} \sqrt{-g }
% = (A^{\mu\nu} \sqrt{-g })_{,\mu} $ I THOUGHT THIS WAS ONLY TRUE FOR
% ANTI-SYMMETRIC TENSORS ${A^{\mu\nu}}={A^{\nu\mu}}$. AM I WRONG?

Recall now Noether's theorem discussed above, the variation of the
Lagrangian may be independently obtained as $\delta (R \sqrt{-g})=( R
\sqrt{-g } \, \delta x^\mu)_{, \mu}$ where $\xi^\mu=\delta x^\mu$.
Therefore, we may consider
\begin{eqnarray}
 0 &=& \delta (R \sqrt{-g}) - ( R \sqrt{-g } \,
\delta x^\mu)_{, \mu} \nonumber  \\
    &=& ( R \xi^\mu \sqrt{-g } )_{, \mu}  
+ 2 ( {\xi^{[\mu ; \nu]}}_{;\mu } \sqrt{-g } )_{,\nu}
+ 2 (R_{\mu \nu}-\frac{1}{2}g_{\mu \nu}R )^{ ; \nu} \sqrt{-g } \xi^\mu
- ( R \sqrt{-g } \, \xi^\mu)_{, \mu} \nonumber  \\
    &=& 2( {\xi^{[\mu ; \nu]}}_{;\mu } \sqrt{-g } )_{,\nu}
+ 2 (R_{\mu \nu}-\frac{1}{2}g_{\mu \nu}R )^{ ; \nu} \sqrt{-g } \xi^\mu  .
\label{VarR0}
\end{eqnarray}

\end{widetext}

Note, that up to this point we have still {\it not\/} used the equations
of motion of general relativity to obtain our Noether current. As noted
in the previous section on the general formulation of Noether currents,
their conservation generically requires imposing the equations of
motion. Indeed, this was presumed to be the case in early work on the
gravitational Noether current.\cite{Iyer1994Ap,Bak1994Ap} Later it was
shown that {\it in the absence of matter\/} the conservation of the
gravitational Noether current does not require the field
equations\cite{Kim2013} (we extend this result to the case including
matter in the following section). In particular, we can see that the
vanishing of the second extra term in Eq.~(\ref{VarR0}) occurs due to
purely geometric behavior. Recall that the definition of the Reimann
tensor naturally implies the Bianchi identities, which may be
written\cite{Poisson2004Ap}
\begin{equation}
R_{\mu\nu\alpha\beta;\gamma} + R_{\mu\nu \gamma \alpha ;\beta}
+ R_{\mu\nu \beta \gamma ; \alpha} = 0.
\label{Bianch1}
\end{equation}
Contracting Eq.~(\ref{Bianch1}) with $g^{\mu \alpha}$ yields
\begin{equation}
R_{\nu \beta;\gamma} - R_{\nu \gamma;\beta}
+ {R_{\mu\nu \beta \gamma}}^{ ; \mu} = 0 ,
\label{Bianch2}
\end{equation}
and further contracting Eq.~(\ref{Bianch2}) with $g^{\nu \gamma}$ yields
\begin{equation}
{R_{\nu \beta}}^{;\nu} - R_{;\beta} + {R_{\mu \beta}}^{ ; \mu} = 0 .
\label{Bianch3}
\end{equation}
Or equivalently,
\begin{equation}
(R_{\mu \nu}-\frac{1}{2}g_{\mu \nu}R )^{ ; \nu} = 0 .
\label{Bianch4}
\end{equation}
Therefore, Eq.~(\ref{VarR0}) may finally be written
\begin{eqnarray}
 0 &=& ( {\xi^{[\mu ; \nu]}}_{;\mu } \sqrt{-g } )_{,\nu} \nonumber  \\
    &=&  {\xi^{[\mu ; \nu]}}_{;\mu \nu} \sqrt{-g }  \nonumber  \\
    &=& {J^\nu}_{;\nu} \sqrt{-g }  \; ,
\end{eqnarray}
and the corresponding Noether current is
\begin{equation}
J^\nu = {\xi^{[\mu ; \nu]}}_{;\mu} .
\label{NoCL}
\end{equation}
We conclude that the detailed Lagrangian formulation of Noether's
theorem agrees with the result from the Hamiltonian formulation
that Eq.~(\ref{NoCL}) is an off-shell Noether current which relies
solely on the geometry of spacetime in the case of pure gravity
(i.e., no matter).\cite{Kim2013}

\section*{Influence of matter fields}

We shall now show that the off-shell nature of the Noether current
conservation remains true even in the presence of matter.
Recall that the Lagrangian of a matter field may be written
${\cal L}_\text{M} \sqrt{-g}$. The variation of this Lagrangian
may then be calculated on the one hand via
\begin{eqnarray}
&&\delta ( {\cal L}_\text{M} \sqrt{-g} ) \nonumber \\ 
&=& \frac{\partial {\cal L}_\text{M}}{\partial x^\mu} \delta x^\mu \sqrt{-g}
+ \frac{\partial {\cal L}_\text{M}}{\partial g^{\mu \nu}}
\delta g^{\mu\nu} \sqrt{-g} - \frac{1}{2} {\cal L}_\text{M}
g_{\mu \nu} \delta g^{\mu \nu} \sqrt{-g}    \nonumber \\
&=&\frac{\partial {\cal L}_\text{M}}{\partial x^\mu} \delta x^\mu \sqrt{-g}
+ \frac{\partial {\cal L}_\text{M}}{\partial g^{\mu \nu}}
\delta g^{\mu\nu} \sqrt{-g} + \frac{1}{2} {\cal L}_\text{M}
g_{\mu \nu} h^{\mu \nu} \sqrt{-g}    \nonumber \\
&=&\frac{\partial {\cal L}_\text{M}}{\partial x^\mu} \delta x^\mu \sqrt{-g}
+ \frac{\partial {\cal L}_\text{M}}{\partial g^{\mu \nu}}
\delta g^{\mu\nu} \sqrt{-g} + {\cal L}_\text{M} {\xi^\mu}_{; \mu} \sqrt{-g} , 
\label{L1}
\end{eqnarray} 
where we have used $h^{\mu \nu} = 2 \xi^{(\mu ;\nu)}$ in the last step.

On the other hand, the variation of ${\cal L}_\text{M} \sqrt{-g}$
may be equally written as
\begin{eqnarray}
\delta ( {\cal L}_\text{M} \sqrt{-g} ) 
&=& ( {\cal L}_\text{M} \sqrt{-g} \xi^\mu )_{,\mu} \nonumber \\
&=& ( {\cal L}_\text{M} \xi^\mu )_{;\mu} \sqrt{-g} \nonumber \\
&=& ( {\cal L}_\text{M} )_{; \mu} \xi^\mu \sqrt{-g}
+ {\cal L}_\text{M} {\xi^\mu }_{;\mu} \sqrt{-g} .
\label{L2}
\end{eqnarray}

Combining the results of Eqs.~(\ref{L1}) and (\ref{L2}) yields
\begin{eqnarray}
&& \delta ( {\cal L}_\text{M} \sqrt{-g} )
- \delta ( {\cal L}_\text{M} \sqrt{-g} ) \nonumber \\
&=& \Bigl(\frac{\partial {\cal L}_\text{M}}{\partial x^\mu} \delta x^\mu
+ \frac{\partial {\cal L}_\text{M}}{\partial g^{\mu \nu}}
\delta g^{\mu\nu} \Bigr) \sqrt{-g} 
% &&\frac{\partial {\cal L}_\text{M}}{\partial g^{\mu \nu}}
% \delta g^{\mu\nu} \sqrt{-g}
- ( {\cal L}_\text{M} )_{; \mu}
\xi^\mu \sqrt{-g} \nonumber \\
&=&\Bigl(\frac{\partial {\cal L}_\text{M}}{\partial x^\mu} \delta x^\mu
+ \frac{\partial {\cal L}_\text{M}}{\partial g^{\mu \nu}}
\delta g^{\mu\nu} \Bigr) \sqrt{-g} - \delta {\cal L}_\text{M} \sqrt{-g}
\nonumber \\
&=& \delta {\cal L}_\text{M} \sqrt{-g}
- \delta {\cal L}_\text{M} \sqrt{-g} \nonumber \\
&\equiv& 0 .
\end{eqnarray}
Here the difference of the variations of
$\delta ( {\cal L}_\text{M} \sqrt{-g} )$ based on the two different
approaches to computing it vanishes. Hence no contribution is made
to the gravitational Noether current from the matter-part of the Lagrangian.

We therefore conclude that the off-shell character of the gravitational
Noether charge appears to be a general result, applying even in the
presence of matter fields. Of course this result appears trivial when
the Hamiltonian formalism for Noether charges is used, but as we have seen
a detailed analysis using the Lagrangian formalism comes to the same
conclusion.

\section*{Influence of the cosmological constant}

The term in the Lagrangian for the cosmological constant may be written
$\alpha \Lambda \sqrt{-g}$ where $\alpha$ is some constant. The 
variation of this may therefore be written on the one hand as
\begin{eqnarray}
\delta ( \alpha \Lambda \sqrt{-g} )
&=& \frac{1}{2} \alpha \Lambda g_{\mu\nu} h^{\mu\nu} \sqrt{-g}  \nonumber \\
&=& \alpha \Lambda g_{\mu\nu} \xi^{\mu ;\nu} \sqrt{-g} .
\end{eqnarray}

On the other hand, this variation may be equally written
\begin{eqnarray}
\delta ( \alpha \Lambda \sqrt{-g} )
&=& \delta ( \alpha \Lambda \sqrt{-g} \xi^\mu)_{,\mu} \nonumber \\
&=& \delta (  \alpha \Lambda  \xi^\mu)_{;\mu} \sqrt{-g} \nonumber \\
&=& \alpha \Lambda {\xi^\mu}_{;\mu} \sqrt{-g} ,
\end{eqnarray}
where we have used $\alpha \Lambda$ is a constant in the last step.
The the difference between these two variations then yields
\begin{eqnarray}
&&\delta ( \alpha \Lambda \sqrt{-g} ) - \delta ( \alpha \Lambda \sqrt{-g} )
\nonumber\\
&=& \alpha \Lambda \sqrt{-g} ( g_{\mu\nu} \xi^{\mu ;\nu} - {\xi^\mu}_{;\mu} )
 \equiv 0.
\end{eqnarray}
Therefore, the cosmological constant term in the Einstein-Hilbert
action density yields a vanishing contribution to the Noether current.

\section*{Integrated conservation laws}

\begin{figure}[ht]
\centering
\includegraphics[width=0.18\textwidth]{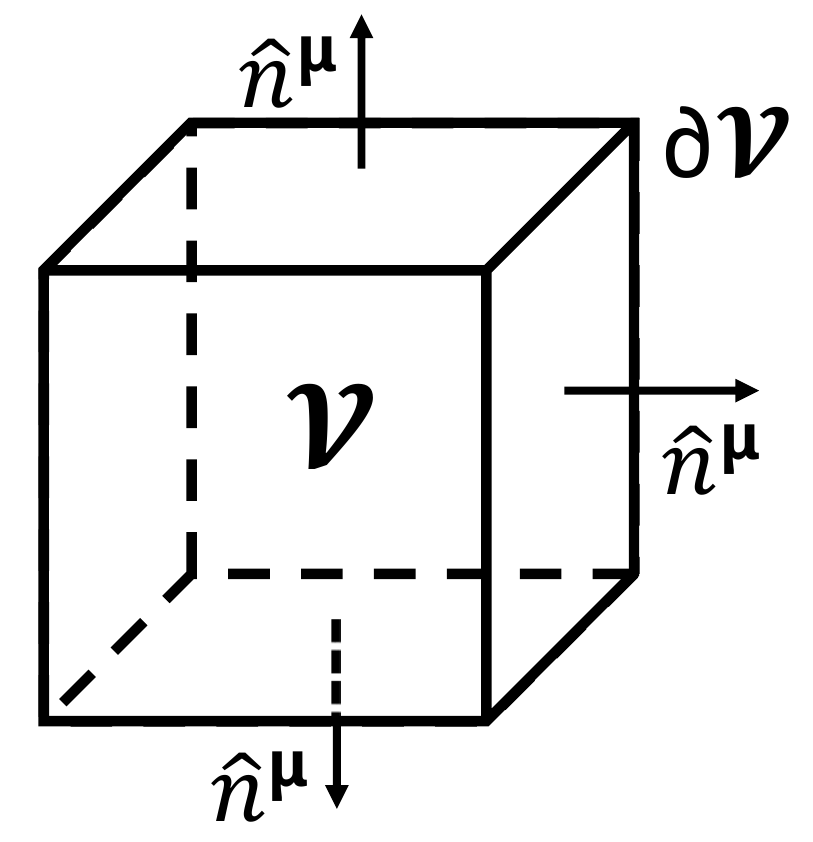}
% \vskip 0.1truein
\caption{This 4-volume $ {\cal V} $ is a subset of the entire spacetime
manifold ${\cal M}$.  Here $ \partial {\cal V} $ is the boundary of $
{\cal V}$, and $\hat n^\mu$ is the outgoing unit vector normal to the
boundary $ \partial {\cal V}$. } 
\label{fig1}
\end{figure}

After proving the generalized Komar current is locally conserved, we now
study the corresponding integrated conservation laws of this current.
Integrating Eq.~(\ref{con0}) (or Eq.~(\ref{NoCL})) over a 4-volume, a
subvolume, $ {\cal V} \subset {\cal M} $ of the entire manifold, yields
\begin{equation}
 \int_{\cal V}   {J^\mu}_{;\mu} \sqrt{-g} \, d^4z = 0,
\end{equation}
and applying Stokes' theorem we find
\begin{equation}
\int_{\partial {\cal V}}   J^\mu \hat
n_\mu \sqrt{\gamma^{({\partial {\cal V}})}}\,  d^3x
= \int_{\partial {\cal V}}  {\xi^{[\nu;\mu]}}_{;\nu}
\hat n_\mu \sqrt{\gamma^{({\partial {\cal V}})}}\,  d^3x = 0 ,
 \label{inte4}
\end{equation}
where $\partial {\cal V}$ is the boundary of ${\cal V}$ and $\hat n^\mu$
is the unit vector normal to ${\partial {\cal V}}$, see Fig.~\ref{fig1},
and $\gamma^{({\cal S})}$ is the determinant of the induced metric on
(sub-)manifold ${\cal S}$. This means that the current flux into the
4-volume is the same as the current flux out. This is a {\it local\/}
conservation law for an {\it arbitrary\/} vector field in an arbitrary
dynamical spacetime.

% \FloatBarrier

\begin{figure}[ht]
\centering
\includegraphics[width=0.25\textwidth]{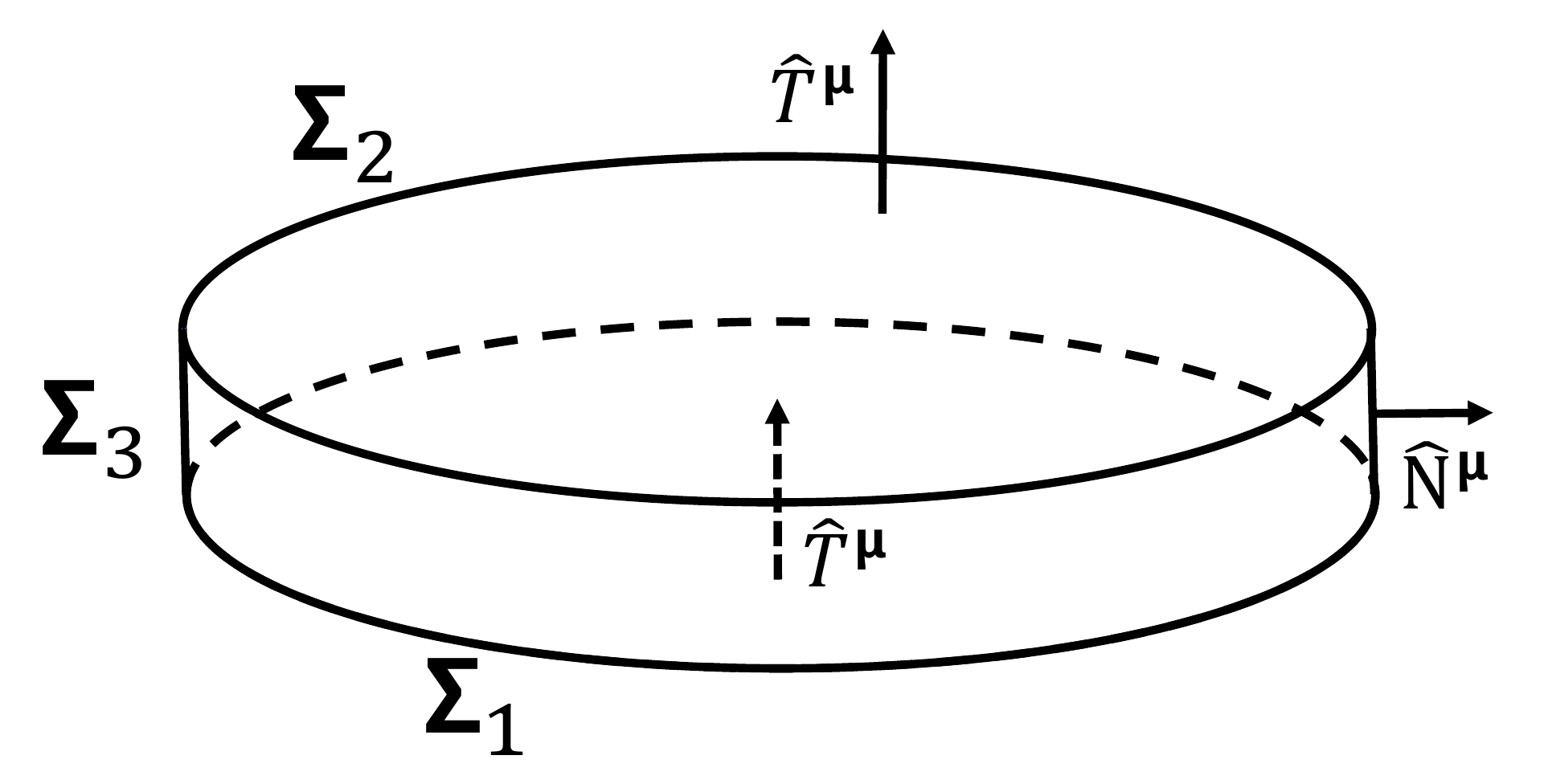}
% \vskip 0.1truein
\caption{This 4-volume $ {\cal V} $ is a region between two three
hypersurface $\Sigma_1$, $\Sigma_2$, and $\Sigma_1$, $\Sigma_2$ and
$\Sigma_3$ together make up of its boundary. Here $\hat T^\mu$ is the
timelike unit normal vector pointing to the future, and $\hat N^\mu$ is
the spacelike outgoing unit vector normal to the side timelike boundary.} 
\label{fig2}
\end{figure}

Next, if we use the 3+1 split method \cite{gourgoulhon2012Ap} to foliate
the spacetime into a family of non-intersecting spacelike hypersurfaces
labeled by $f$ with net flux $J^{\mu} \hat N_\mu$ out through the side
timelike boundary $\Sigma_3$ ($\hat N^\mu$ is the spacelike unit vector
normal to the boundary). Consider a volume ${\cal V}$ consisting of the
region between a pair of such hypersurfaces $\Sigma_1, \Sigma_2$ and
side boundary $\Sigma_3$ (see Fig.~\ref{fig2}) then from
Eq.~(\ref{inte4}) we find
\begin{eqnarray}
    &&\int_{\Sigma_1} {\xi^{[\nu;\mu]}}_{;\nu}
\hat T_\mu \sqrt{\gamma^{(\Sigma_1)}} \, d^3x  \nonumber \\
    &=& \int_{\Sigma_2} {\xi^{[\nu;\mu]}}_{;\nu}
\hat T_\mu \sqrt{\gamma^{(\Sigma_2)}} \, d^3x
+  \int_{\Sigma_3} {\xi^{[\nu;\mu]}}_{;\nu}
\hat N_\mu \sqrt{\gamma^{(\Sigma_3)}} \, d^3x , \nonumber \\
 \label{inte41}
\end{eqnarray}
where $\hat T_\mu$ is the future directed timelike unit normal to the
hypersurfaces. Note that this split formalism is unnecessary to have any
connection with the coordinates system although using the coordinate to
label the hypersurface sometimes may simplify the calculation.  When the
Noether current (some kind of 'energy' current) $J^\mu \hat N_\mu$
vanishes on the side boundary $\Sigma_3$, this integral over the
three-dimensional hypersurface $\Sigma_i$ will be independent of the
hypersurface.

% \FloatBarrier

Therefore, on arbitrary spacelike three hypersurface, we can always
have such a well-defined Noether current
\begin{eqnarray}
  \int_{\Sigma}
{\xi^{[\nu;\mu]}}_{;\nu} \hat T_\mu \sqrt{\gamma^{(\Sigma)}}\,  d^3x , 
 \label{inte42}
\end{eqnarray}

Recalling Stokes's theorem for an
anti-symmetric tensor\cite{carroll2004Ap} $F^{\mu\nu}$ 
\begin{equation}
    \int_\Sigma \hat T_\mu {F^{\mu \nu}}_{;\nu}
\sqrt{\gamma^ {(\Sigma)} } d^{n-1}x =\! \int_{\partial \Sigma}\!
\hat T_\mu F^{\mu \nu} \hat N_\nu \sqrt{\gamma^{(\partial \Sigma)} }
d^{n-2}y,
    \label{stokes}
\end{equation}
where $\hat N^\mu$ is the outgoing spacelike unit vector normal to
$\partial \Sigma$.
Applying this Stokes's theorem to Eq.~(\ref{inte42}) yields
\begin{eqnarray}
  \int_{\partial \Sigma } \!
\xi^{[\nu;\mu]} \hat N_\nu \hat T_\mu 
    \sqrt{\gamma^{(\partial \Sigma)}   } \,d^2y = \int_{\partial \Sigma } \!
\xi^{[\nu;\mu]} d\Sigma_{\mu\nu}, 
 \label{inte43}
\end{eqnarray}
where $d\Sigma_{\mu\nu} = \hat N_{[\nu} \hat T_{\mu]}
\sqrt{\gamma^{(\partial \Sigma)} } \,d^2y
= \hat N_{[\nu} \hat T_{\mu]} dA$.

\section*{Proof that $\kappa_\text{Noether}=\kappa_\text{Killing}$
on Killing horizons}

Recall that in the manuscript, we define a natural surface gravity for
each point on the world tube ${\cal W}$, rescaled to the rate-of-change
of coordinate time, via
\begin{equation}
\kappa_{\text{Noether}} \equiv \Phi \,a_\mu \hat R^\mu,
\label{kappaNAp}
\end{equation}
where $a_\mu\equiv v_{\mu;\nu} v^\nu$ is the 4-acceleration of an
observer from our congruence passing though this point, and $\hat R^\mu$
is normal to ${\cal W}$ at this point.

Now let us consider a non-degenerate Killing
horizon with Killing vector
\begin{equation}
K^\mu = \partial_t^\mu + \Omega_\text{H}\, \partial _\phi^\mu,
\end{equation}
where $t$ is the coordinate time at spatial infinity, $\phi$ is the
azimuthal angular coordinate, and the constant $\Omega_\text{H}$ is the
angular velocity of the horizon for a Kerr black hole. According to the
conventional definition, the surface gravity, $\kappa_\text{Killing}$,
for such a Killing horizon satisfies
\begin{equation}
{K^\mu}_{;\nu} K^\nu = \kappa_\text{Killing} \, K^\mu ,
\label{kappaKAp}
\end{equation}
on the horizon.

\vskip 0.1in

\noindent
{\bf Lemma 1:} The Noether surface gravity, $\kappa_\text{Noether}$,
reduces to the standard result for Killing horizons,
$\kappa_\text{Killing}$, for non-degenerate, non-bifurcate Killing
horizons.

\vskip 0.1in
\noindent
{\bf Proof:} We start by noting that the definition of Eq.~(\ref{kappaKAp})
is not suitable to be treated as a limit, since the two sides of this
equation can only be parallel for $K^\mu$ null. Instead, we use an
alternative characterization of this surface gravity.\cite{Wald1984Ap} In
order to formulate this alternative, consider a timelike `Killing
observer' with 4-velocity
\begin{equation}
v^\mu \equiv K^\mu / \sqrt{-K^\mu K_\mu},
\end{equation}
situated outside the horizon. After a proper time, $\delta \tau$, such
an observer will have moved by
\begin{equation}
\delta x^\mu = v^\mu \delta \tau.
\end{equation}
Taking the inner product of both sides of this equation with
$\nabla_\mu t$ yields
\begin{equation}
\delta t = \delta \tau / \sqrt{-K^\mu K_\mu}.
\end{equation}
Thus the red-shift factor associated with this observer satisfies
$\Phi^2 = -K^\mu K_\mu$. We are now in a position to give the alternate
formulation for the surface gravity as\cite{Wald1984Ap}
\begin{equation}
\kappa_\text{Killing} = \text{lim} \;\Phi\, a,
\label{lim}
\end{equation}
where $a^2=a^\mu a_\mu$ is the square of the 4-length of the
4-acceleration $a^\mu=v^\mu_{~~;\nu} v^\nu$ and the limit corresponds to
considering Killing observers ever closer to the horizon.

We note that in the limit of approaching the horizon $K^\mu$ is tangent
to the Killing horizon world tube, so the Killing observers studied
above may be considered as a suitable congruence of observers for the
purposes of our Noether charge surface gravity of Eq.~(\ref{kappaNAp}).
Recalling that $\Phi^2 = -K^\mu K_\mu=0$ on the Killing horizon.
It follows that derivatives of any non-trivial function of $\Phi$
must therefore be normal to the horizon world tube, along $\hat R^\mu$. We
may now write
\begin{eqnarray}
-(K^\mu K_\mu)_{;\nu} &\propto & \hat R_\nu \nonumber \\
\Rightarrow ~~
 -K_{\mu ; \nu} K^\mu  &\propto & \hat R_\nu
\nonumber \\
\Rightarrow  ~~~~~\,
 K_{\nu ; \mu} K^\mu  &\propto & \hat R_\nu \nonumber \\
\Rightarrow ~~~~~~~~~~~~~
 a_\nu  &\propto & \hat R_\nu ,
\end{eqnarray}
where the Killing condition $K_{\mu ; \nu} = -K_{\nu ; \mu}$ is used
in the third line. It follows that $a=a_\mu \hat R^\mu$ and we find
that for non-degenerate non-bifurcate Killing horizons that
\begin{equation}
\kappa_{\text{Noether}}=\kappa_{\text{Killing}}.
\end{equation}
If the horizons are either degenerate or bifurcate the conventional
definition from Eq.~(\ref{kappaKAp}) yields $\kappa_{\text{Killing}}\equiv
0$, whereas neither the alternative formulation of Eq.~(\ref{lim}) for
Killing horizons nor our Noether charge surface gravity of
Eq.~(\ref{kappaNAp}) suffer from this unphysical behavior. \qed

\section*{Local Lorentz transformation of the redshift}

\begin{figure}[ht]
\centering
\vskip -0.1truein
\includegraphics[width=0.48\textwidth]{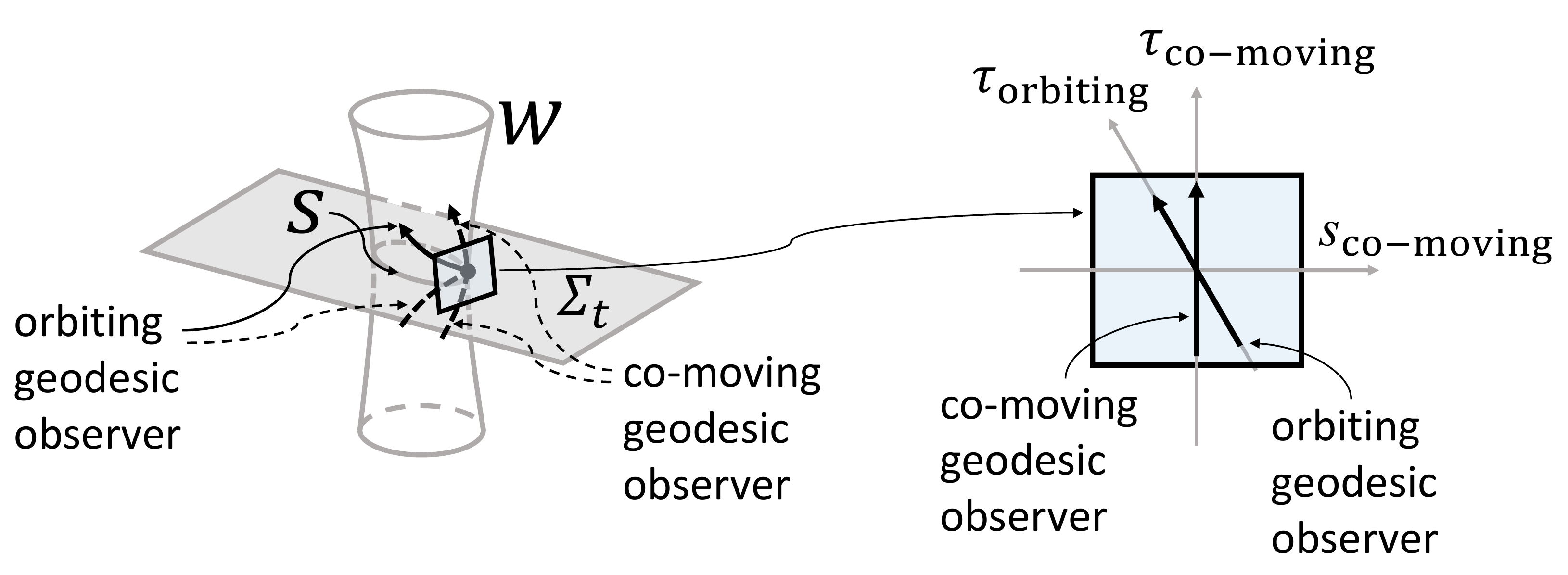}
\vskip -0.1truein
\caption{On the left is Figure~2 with its non-geodesic observer moving
tangent to the world tube ${\cal W}$; the instantaneously co-moving
geodesic observer in a parabolic-like trajectory; and the orbiting
geodesic observer, in this example, spiraling around ${\cal W}$. A
small patch around the intersection point is expanded on the right. This
shows the co-moving and orbiting observers' world lines on a flat
spacetime diagram. This shows that the relation between the red shift
factors for these observers is determined by a local Lorentz
transformation.}
\vskip -0.1truein
\label{Transformation}
\end{figure}

We now consider how the red-shift factor for a non-geodesic, (or the
co-moving) observer may be obtained by observation of the redshift of an
orbiting (geodesic) observer (see Fig.~\ref{Transformation}).

The key point is that we assume that this latter red shift factor,
$\Phi_\text{orbiting}$, is known through observation. To effect the
local Lorentz transformation we also need to know the velocity, $v$, of
these latter two observers relative to each other. In special relativity
the velocities of either observer relative to the other are identical
(up to a minus sign) even though measured in very different ways
\begin{equation}
v=\frac{ds_\text{orbiting}}{d\tau_\text{orbiting}}
=-\frac{ds_\text{co-moving}}{d\tau_\text{co-moving}},
\label{vrel}
\end{equation}
where $s_{i}$ is the (proper length) location of observer $j\ne i$
in the frame of observer $i$, $\tau_{i}$ is that observer's
proper time and $i,j\in\{\text{orbiting},\text{co-moving}\}$.

For simplicity, we assume that the reference observer, who is making the
astronomical observation, is in the instantaneous rest frame of the
non-geodesic observer who is accelerating to remain stationary on the
world tube, we therefore use the latter expression of Eq.~(\ref{vrel}).
The time-component of the Lorentz transformation then
yields %\cite{Landau1975Ap}
\begin{equation}
\Phi_\text{non-geodesic}=\Phi_\text{co-moving}
=\frac{1}{\sqrt{1-v^2}}\, \Phi_\text{orbiting}.
\label{PhiPhi}
\end{equation}

Let us consider this in a simple example involving circular equatorial
orbits in the Schwarzschild metric
\begin{equation}
ds^2 = - \Bigl( 1-\frac{2 M}{r} \Bigr) dt^2
+ \Bigl(1-\frac{2 M}{r}\Bigr)^{\!-1}\! dr^2 + r^2 d\Omega^2 .
\label{metricS}
\end{equation}
Assuming orbits of the form $v^\mu\propto (1,0,0,f(r))$, and
using the conditions
${v^\mu}_{;\nu} v^\nu = 0$ and $v^\mu v_\mu=-1$, we find
\begin{equation}
v^\mu = \frac{dx_\text{orbiting}^\mu} {d\tau_\text{orbiting} }
= \frac{1}{\sqrt{1-\frac{3 M}{r}}}
\biggl( 1,0,0,\frac{1}{r} \sqrt{\frac{M}{r}}\, \biggr) .
\end{equation}
From Observation~1, the orbital red shift factor is
\begin{equation}
\Phi_\text{orbiting} = \sqrt{1-{{\frac{3 M}{r}}}},
\label{Phio}
\end{equation}
which we assume to be observationally accessible.

With regard to coordinate time, the orbit follows
\begin{equation}
\frac{dx_\text{orbiting}^\mu} {dt }
=  ( 1, 0, 0, \frac{1}{r}\sqrt{\frac{M}{r}}) ,
\end{equation}
Thus, in the rest frame the velocity is
$ds_\text{co-moving}/dt = r d\phi/ dt = \sqrt{M/r} $,  still
measured with-regard-to coordinate time.
In terms of the proper time for the co-moving observer
\begin{equation}
v \equiv \frac{ds}{d \tau_\text{co-moving}}
=  \frac{1}{\Phi_\text{co-moving}}\, \sqrt{\frac{M}{r}} ,
\end{equation}

From Eq.~(\ref{PhiPhi}) we then obtain
\begin{eqnarray}
\Rightarrow ~~~~~~~~~~~~~~~~ \Phi_\text{co-moving}^2 (1-v^2)
&=& \Phi_\text{orbiting}^2 \nonumber \\
\Rightarrow ~~ \Phi_\text{co-moving}^2
\Bigl( 1- \frac{M}{r \, \Phi_\text{co-moving}^2} \Bigr)
&=& 1-\frac{3 M}{r} \nonumber \\
\Rightarrow ~~~~~~~~~~~~~~~~~~~~~~~~~~ \Phi_\text{co-moving}^2
&=& 1-\frac{2 M}{r} \nonumber \\
\Rightarrow ~~~~~~~~~~~~~~~~~~~~~~~~~~ \Phi_\text{co-moving}
&=& \sqrt{1-\frac{2 M}{r} }  ,
\end{eqnarray}
where we used Eq.~(\ref{Phio}) to obtain the second line.

Note that we only used the Lorentz transformation summarized in
Eq.~(\ref{PhiPhi}), and in principle observationally accessible
values for $\Phi_\text{orbital}$ and the orbital velocity $v$.
Of course, direct access to the metric, Eq.~(\ref{metricS}) would
have come to the same conclusion, but this is not directly observable.

As a final observation, we note that given a sufficient number of
similar observations at differing locations, we would have been able to
determine, for example, that 
\begin{equation}
M_\text{Grav} = M,
\end{equation}
all concentrated within a central region within $r< 3M$ (assuming
access to circular orbits only).

\section*{APPENDIX 2: RELATED LITERATURE}

\section*{Brown-York quasi-local mass}

In 1993, Brown and York proposed a quasi-local mass/energy
definition.\cite{Brown1993Ap} Their proposal is based on the 3+1 split
formalism and assumes the 3-dimensional boundary (that we call the
observers' world tube) is orthogonal with the 3-dimensional foliated
hypersurfaces. Thus in their construction, $v^\mu=\hat T^\mu$ and $\hat
R^\mu = \hat N^\mu$. Therefore, the projector onto the two-surface
${\cal S}$ may be written $\sigma_{\mu \nu}= g_{\mu \nu} + \hat T_\mu
\hat T_\nu - \hat N_\mu \hat N_\nu$. The extrinsic curvature of the
two surface ${\cal S}$ is hence given by $k=\hat N_{\mu ;
\nu} \sigma^{\mu \nu}$ with respect to the hypersurface that ${\cal S}$ is
embedded in. Finally, Brown and York define a quasi-local mass
\begin{equation}
M_\text{B-Y} = - \frac{1}{8 \pi}\int_{\cal S} (k-k_0)\, dA,
\end{equation}
where $k_0$ is the extrinsic curvature of ${\cal S}$ on a background
flat spacetime.

Although Brown and York claim this proposal to be a quasi-local mass, it
seems it only works at spatial infinity as a global mass
definition.\cite{Poisson2004Ap} In particular, though it reduces to the
ADM mass at infinity for asymptotically-flat spacetimes, it does not
yield the standard results over finite surfaces, even for a spacetime
with a single black hole. For the Schwarzschild metric, it is easy to
compute $k=\frac{2}{r} \sqrt{1-\frac{2 M}{r}}$ and $k_0=\frac{2}{r}$ by
taking $M\rightarrow 0$. Thus
\begin{eqnarray}
M_\text{B-Y} &=& - \frac{1}{8 \pi}\int (k-k_0)\, dA \nonumber \\
&=& - \frac{1}{8 \pi}\int \Big[-\frac{2M}{r^2}
- \frac{M^2}{r^3} + O\Big(\frac{1}{r^4}\Big) \Big] dA ,\nonumber\\
&=& M+\frac{M^2}{2r}+O\Big(\frac{1}{r^2}\Big).
\end{eqnarray}
Therefore the Brown-York mass only yields the standard result when $r
\rightarrow  \infty$. Indeed, some textbooks only describe this mass
definition as applying at spatial infinity, see
Ref.~[\onlinecite{Poisson2004Ap}].

\section*{Alternative dynamical surface gravities}

Traditionally, surface gravity is defined for stationary spacetime on a
Killing horizon. However, a real physical black hole in our Universe
must interact with other gravitating body and hence be dynamical. So
several different surface gravity definitions for dynamical spacetimes
have been proposed and they do not agree with each other even for the
most simple spherically-symmetric dynamical scenarios.\cite{Nielsen2008Ap}
We give a short review to some of these definitions here.

The first definition for a dynamical, non-Killing horizon, surface
gravity proposed by Hayward,\cite{Hayward1994Ap} is independent of the
chosen normalization on the horizon
$\kappa^\text{H}=\frac{1}{2}\sqrt{-\theta^{(l)}_{\;\;\; ;\mu}n^\mu }$.
However, it is known that this does not give the correct answer in the
Reissner–Nordstrom case even for spherically symmetric spacetimes.

Since the Killing vector used in the traditional surface gravity
definition is null on the Killing horizon, a generalized surface gravity
may be proposed based on the outward null vector $l^\mu$ that
${l^\mu}_{; \nu} l^\nu = \kappa^\text{null} l^\mu$. However, there are
two problems for this definition: (i) There is a freedom in the
normalization of such a null vector and hence the surface gravity is not
uniquely defined. (ii) The horizon tube of a truly dynamical horizon may
not be null and hence this null vector may not be tangent to the horizon
tube (while in Killing horizon $\xi^\mu$ is tangent to the horizon
tube).

To fix the parametrization problem in the surface gravity based on
outward null vector, Fodor et al.\cite{Fodor1996Ap} proposed a non-local
choice of normalization based on the ingoing null geodesic $n^\mu$. They
requires that $l^\mu n_\mu=-1$ and $t^\mu n_\mu = -1$ where $t^\mu$ is
the asymptotically time-translational Killing vector for an
asymptotically-flat spacetime. Then the surface gravity is defined as
$\kappa^\text{F}=-l_{\mu ; \nu}l^\nu n^\mu$. However, this is only
proposed for spherically symmetric spacetimes and requires the spacetime
to be asymptotically flat.

Hayward also proposed a dynamical surface gravity for spherically
symmetric spacetimes in terms of the Kodama vector. However, it is also
only designed for spherically symmetric spacetimes.\cite{Hayward1998Ap} It
seems this only work for some special kind of coordinates system even
for spherically symmetric spacetimes.

For an isolated horizon,\cite{Ashtekar2000Ap} Ashtekar and colleagues also
propose $\kappa^\text{B}= - l_{\mu ; \nu} l^\nu n^\mu$. To solve the
normalization problem, they fix the surface gravity as a unique function
of the horizon area and energy of the black hole, in terms of the known
thermal relation in the static case. However, this is an effective
surface gravity which means it is an average of the real local surface
gravity and it does not is hard to deal with for scenarios with several
horizon parameters, like the Einstein-Yang-Mills case. 

Booth and Fairhurst generalized the suggestion of Ashtekar et al when
they try to generalize the isolated horizon to the so-called slowly
evolving horizon.\cite{Booth2004Ap} They suggest $\kappa^\text{B}= - B
l_{\mu ; \nu} l^\nu n^\mu - C n_{\mu ; \nu} n^\nu l^\mu$ with the normal
of the horizon equals $B l^\mu + C n^\mu$ for the slowly evolving
horizon. The normalization of this surface gravity also needs the help
of the horizon parameters and the hold of the first law of black hole
thermodynamics. So this definition is not self-consistent by itself too.

For dynamical horizon,\cite{Ashtekar2004Ap} an effective surface gravity
$\kappa=\frac{1}{2r}\frac{d f(r)}{ dr}$ is used in the study of black
hole thermodynamics by Ashtekar and Krishnan. Once again there is
freedom in the normalization that can usually be fixed by appeal to the
stationary Kerr solution. Moreover, since it is derived based from an
area balanced law, it more like an average of the real surface gravity
for truly dynamical system.

For these proposed surface gravity definitions for dynamical spacetimes,
only $\kappa^\text{F}$ perfect agrees with the results calculated by the
conventional surface gravity,\cite{Nielsen2008Ap} but it only defines for
spherically symmetric scenarios. All the other proposals either need
some specific normalization or only works for some special coordinates
system even for the spherically symmetric spacetimes.

% \vskip 0.1in

\section*{Entropy as a Noether charge?}

In 1993, Wald\cite{Wald1993Ap,Iyer1994Ap} considered generally
diffeomorphic theories of gravity and found a `first law' for black
holes perturbed form stationarity of the form
\begin{equation}
\delta \!\!\int\! Q = \delta {\cal E} - \text{angular momentum terms},
\label{first}
\end{equation}
where the integral is taken over the horizon surface (a 2-surface in
3+1 dimensions) and 
where $\delta$ denotes a diffeomorphic perturbation, $Q$ is the Noether
charge for the theory (here being integrated over the black hole's
horizon), and ${\cal E}$ is the spacetime's ADM energy. He argued that
this corresponded precisely to the first law of black hole mechanics
from which he concludes
\begin{equation}
\delta \!\!\int\! Q = \frac{\kappa}{2\pi}\, \delta S,
\label{dQ}
\end{equation}
where $\kappa$ is the unperturbed Killing surface gravity and
$S$ the presumed black hole entropy which he identifies as
\begin{equation}
S \equiv \frac{2\pi}{\kappa}\!\int Q .
\label{S}
\end{equation}
(Wald\cite{Wald1993Ap} writes this as $S=2\pi\! \int^{\strut\text{$~$}}\!\!\!
\tilde Q$, in terms of the Noether charge, $\tilde Q=Q/\kappa$, obtained
from a Killing field normalized to have unit surface gravity, however,
this is simply Eq.~(\ref{S}). It is in this normalized form that Wald
calls the entropy as an integrated Noether charge.\cite{Wald1993Ap}) 

A problem immediately arises from this analysis if one combines
Eqs.~(\ref{dQ}) and~(\ref{S}) to yield
\begin{equation}
\delta (\kappa S) = \kappa\, \delta S.
\end{equation}
It is easy to see that this relation fails to hold, for example,
reducing to $1=2$ for Schwarzschild black holes in ordinary general
relativity, under a diffeomorphism that infinitesimally changes a black
hole's mass. This leads to the likely conclusion that Eq.~(\ref{first})
was incorrectly identified as the first law of black hole mechanics.

Indeed, Wald's Eq.~(\ref{S}) has received support for equilibrium
black holes. For example, Garfinkle and Mann\cite{Garfinkle2000Ap}
consider the so-called generalized gravitational
entropy,\cite{Hawking1999aAp,Hawking1999bAp} $S_{\text{gen}}$, in the
Euclidean domain. Adding terms to the gravitational action which
preserve the equations of motion they find
\begin{equation}
S_{\text{gen}} = \beta\, \Big(\int\! Q + \int_\infty \! Q_0\biggr),
\end{equation}
where the temperature $1/\beta$ is given by the periodicity
of Euclidean time and the final integral denotes evaluating $Q$ on a
background spacetime at spatial infinity.

\vskip 0.1in

Wald's original analysis\cite{Wald1993Ap} was generalized in
Ref.~[\onlinecite{Iyer1994Ap}] with the entropy defined as an integral
of only part of the Noether current, though still keeping the
claim that Eq.~(\ref{dQ}) is key to a first law of black hole
mechanics. However, as we shall now see, this modification still
leads to an inconsistency when applied to the simplest case of a
Schwarzschild black hole in 3+1 spacetime dimensions.

Following Ref.~[\onlinecite{Iyer1994Ap}] we take $\xi^\mu = \partial_t$
for the Schwarzschild spacetime and the integral of $Q=Q[\xi^\mu]$ at
spatial infinity (denoted as $\infty$) is identified in
Ref.~[\onlinecite{Iyer1994Ap}] as precisely one-half of the expression
for the Komar mass (see Eq.~(85) of that reference), i.e., we must
have
\begin{eqnarray}
\int_\infty Q[\partial_t] = \frac{1}{2} M . \label{Kinf}
\\ \nonumber
\end{eqnarray}

Now in a vacuum spacetime the Komar integral is well known
to be independent of the boundary of integration,\cite{Wald1984Ap} 
so Eq.~(\ref{Kinf}) may be rewritten as an integral over the horizon
2-surface $H$ (labeled $\Sigma$ in Ref.~[\onlinecite{Iyer1994Ap}]) as
\begin{eqnarray}
\int_\infty Q[\partial_t] =
\int_{H} \!\!Q[\partial_t]
&=& \int_{H}\!\! Q[\xi^\mu] = \frac{1}{2} M.
%\\ \nonumber
\end{eqnarray}

Thus, we may explicitly evaluate the left-hand-side of
Eq.~(\ref{dQ}) for a Schwarzschild black hole as
\begin{equation}
\delta\! \int_{H }\!\! Q[\xi^\mu] = \frac{1}{2}\, \delta M
\label{left}.
\end{equation}
Since the horizon entropy and surface gravity of a Schwarzschild black
hole are $S=4\pi M^2$ and $\kappa = 1/(4M)$, respectively, the
right-hand-side of Eq.~(\ref{dQ}) reduces to
\begin{equation}
\frac{\kappa}{2 \pi} \delta S = \frac{1}{8 \pi M}\, 8\pi M \delta M
= \delta M.
\label{right}
\end{equation}
We now easily see that Eq.~(\ref{dQ}) leads to an explicit contradiction
for even the simplest scenario.

In any of the analyses above, and consistent with what we find for the
surface gravity constrained to ordinary general relativity, the
integrated Noether charge is better described as proportional to the
product $\kappa \,S/(2\pi)$, i.e., physically as an energy, instead of
an entropy. We note that Komar\cite{Komar1959Ap} had already made this
observation, although without explicitly recognizing his analysis as being
related to a Noether charge.

\end{document}